\documentclass[12pt,a4paper,dvips]{article}
\usepackage{a4p}
\usepackage{cite,mcite}
\usepackage{graphicx}
\usepackage{physics}
\usepackage{amssymb}
\usepackage{l3_titlePH,ifthen}

\preprint{2004-002}
\date{January 30, 2004}
\journalname{Phys. Lett. B}

\newlength{\capindent}
\newlength{\capwidth}
\newlength{\figwidth}
\newcommand{\icaption}[2][!*!,!]{\hspace*{\capindent}%
  \begin{minipage}{\capwidth}
    \ifthenelse{\equal{#1}{!*!,!}}%
      {\caption{#2}}%
      {\caption[#1]{#2}}
  \end{minipage}}
\def\tanw{\ensuremath{\tan\!\theta_{W}}}%
\def\sintw{\ensuremath{\sin^2\!\theta_{W}}}%
\def\costw{\ensuremath{\cos^2\!\theta_{W}}}%
\def\sinww{\ensuremath{\sin\!2\theta_{W}}}%
\def\cosww{\ensuremath{\cos\!2\theta_{W}}}%
\def\tantw{\ensuremath{\tan^2\!\theta_{W}}}%
\def\thw{\ensuremath{\theta_{W}}}%
\def\WWstar{\ensuremath{\mathrm{W}\mathrm{W^{(\ast)}}}}%
\def\X{\ensuremath{\mathrm{X}}}%
\def\A{\ensuremath{\mathrm{A}}}%
\def\Ho{\ensuremath{\mathrm{H}}}%
\def\em{\ensuremath{\mathrm{e^-}}}%
\def\ep{\ensuremath{\mathrm{e^+}}}%
\def\dg1z{\ensuremath{\Delta g_1^{\Zo}}}%
\def\dkg{\ensuremath{\Delta \kappa_\gamma}}%
\def\d{\ensuremath{d}}%
\def\db{\ensuremath{d_B}}%
\def\WWZ{\ensuremath{\mathrm{W}\mathrm{W}\mathrm{Z}}}%
\def\WWG{\ensuremath{\mathrm{W}\mathrm{W}\gamma}}%
\def\HGG{\Ho\gamma\gamma}%
\def\HZG{\Ho\Zo\gamma}%
\def\HZZ{\Ho\Zo\Zo}%
\def\HWW{\Ho\W\W}%
\def\eehg{\ensuremath{\ee \ra \Ho\gamma}}%
\def\eehz{\ensuremath{\ee \ra \Ho\Zo}}%
\def\eehee{\ensuremath{\ee \ra \ee\Ho}}%
\def\htogg{\Ho\ra\gamma\gamma}%
\def\htobb{\Ho\ra\bbbar}%
\def\htozg{\Ho\ra\Zo\gamma}%
\def\htoww{\Ho\ra\WWstar}%
\def\htoff{\Ho\ra\ffbar}
\def\eeggg{\ensuremath{\ee \ra \Ho\gamma \ra \gamma\gamma\gamma}}%
\def\eebbg{\ensuremath{\ee \ra \Ho\gamma \ra \bbbar\gamma}}%
\def\eeeegg{\ensuremath{\ee \ra \ee\Ho \ra \ee\gamma\gamma}}%
\def\eeeebb{\ensuremath{\ee \ra \ee\Ho \ra \ee\bbbar}}%
\def\eezgg{\ensuremath{\ee \ra \Ho\gamma \ra \Zo\gamma\gamma}}%
\def\eewwg{\ensuremath{\ee \ra \Ho\gamma \ra \WWstar\gamma}}%
\def\eetoeegg{\ensuremath{\ee \ra \ee\gamma\gamma}}%
\def\eetoggg{\ensuremath{\ee \ra \gamma\gamma\gamma}}%
\def\eetozgg{\ensuremath{\ee \ra \Zo\gamma\gamma}}%
\def\eetowwg{\ensuremath{\ee \ra \WWstar\gamma}}%
\def\Brhgg{\ensuremath{\mathrm{Br}(\Ho\ra\gamma\gamma)}}%
\def\Brhzg{\ensuremath{\mathrm{Br}(\Ho\ra\Zo\gamma)}}%
\def\Brhww{\ensuremath{\mathrm{Br}(\Ho\ra\WWstar)}}%
\def\Ghgg{\ensuremath{\Gamma(\Ho\ra \gamma\gamma)}}%
\def\fpfpbar{\antibar{\mathrm{f^\prime}}}%

\begin{document}
\begin{titlepage}
\title{Search for Anomalous Couplings\\
 in the Higgs Sector at LEP}
\author{The L3 Collaboration}
%
%

\begin{abstract}
Anomalous couplings of the Higgs boson are searched for through
the processes $\eehg$, $\eehee$ and $\eehz$. The mass range $70 \GeV <
\MH < 190 \GeV$ is explored using 602\,$\pb$ of integrated luminosity
collected with the L3 detector at LEP at centre-of-mass energies
$\rts=189 -209 \GeV$.  The Higgs decay channels $\htoff$, $\htogg$,
$\htozg$ and $\htoww$ are considered and no evidence is found for
anomalous Higgs production or decay.  Limits on the anomalous
couplings $\d$, $\db$, $\dg1z$, $\dkg$ and $\xi^2$ are derived as well
as limits on the $\Ho\ra\gamma\gamma$ and $\Ho\ra\Zo\gamma$ decay
rates.

\end{abstract}

\submitted

\end{titlepage}

\section{Introduction}

The mechanism of spontaneous symmetry breaking is a cornerstone of the
Standard Model of the electroweak interactions~\cite{standard_model}.
It explains the observed masses of the elementary particles and
postulates an additional particle, the Higgs boson. Despite its
relevance, experimental information on the Higgs boson is scarce and
indirect. It leaves room for deviations from the Standard Model
expectations such as anomalous couplings of the Higgs boson.

The Standard Model can be extended, via a linear representation of the
$SU(2)_L \times U(1)_Y$ symmetry breaking
mechanism~\cite{linear_realization}, to higher orders where new
interactions between the Higgs boson and gauge bosons become
possible. These modify the production mechanisms and decay properties
of the Higgs boson. The relevant CP-invariant Lagrangian terms
are~\cite{eboli_concha}:
\begin{eqnarray}
{\cal L}_{\rm eff} = g_{\HGG}~\Ho \A_{\mu\nu} \A^{\mu\nu}
       ~+~g^{(1)}_{\HZG}~\A_{\mu\nu} \Zo^{\mu} \partial^{\nu} \Ho
       ~+~g^{(2)}_{\HZG}~\Ho \A_{\mu\nu} \Zo^{\mu\nu} \nonumber \\
       ~+~g^{(1)}_{\HZZ}~\Zo_{\mu\nu} \Zo^{\mu} \partial^{\nu} \Ho
       ~+~g^{(2)}_{\HZZ}~\Ho \Zo_{\mu\nu} \Zo^{\mu\nu}
       ~+~g^{(3)}_{\HZZ}~\Ho \Zo_\mu \Zo^\mu \nonumber \\
       ~+~g^{(1)}_{\HWW}~(\W^{+}_{\mu\nu} \W_{-}^\mu \partial^{\nu} \Ho
       ~+~ h.c.)
       ~+~g^{(2)}_{\HWW}~\Ho \W^{+}_{\mu\nu} \W_{-}^{\mu\nu},
       \label{eq:lagrangian}
\end{eqnarray}

\noindent
where $\A_\mu$, $\Zo_\mu$, $\W_\mu$ and $\Ho$ are the photon, $\Zo$,
$\W$ and Higgs fields, respectively, and $\X_{\mu\nu}= \partial_\mu
\X_\nu - \partial_\nu \X_\mu$. The couplings in this Lagrangian
are parametrized as~\cite{gounaris,hagiwara_ww,wudka}:
\begin{eqnarray}
g_{\HGG} & = & \frac{g}{2 \MW}~\left( \d\sintw + \db \costw \right)
              \label{eq:gg} \\
g^{(1)}_{\HZG} & = & \frac{g}{\MW}~\left(\dg1z \sinww -
                                           \dkg \tanw \right) \\
g^{(2)}_{\HZG} & = & \frac{g}{2 \MW}~\sinww~\left( \d - \db \right)
              \label{eq:zg} \\
g^{(1)}_{\HZZ} & = & \frac{g}{\MW}~\left(\dg1z \cosww +
                                           \dkg \tantw \right)  \\
g^{(2)}_{\HZZ} & = & \frac{g}{2 \MW}~\left( \d\costw + \db \sintw \right) \\
g^{(3)}_{\HZZ} & = & \frac{g~\MW}{2~\costw} \delta_\Zo \\
g^{(1)}_{\HWW} & = & \frac{g~\MW}{\MZ^2} \dg1z  \\
g^{(2)}_{\HWW} & = & \frac{g}{\MW} \frac{\d}{\cosww}, 
\end{eqnarray}

\noindent where $g$ is the $ SU(2)_L$ coupling constant, $\thw$ is
the weak mixing angle and $\MW$ and $\MZ$ represent the masses of
the $\W$ and $\Zo$ bosons, respectively. The five dimensionless
parameters $\d$, $\db$, $\dg1z$, $\dkg$ and $\delta_\Zo$
constitute a convenient set to describe deviations in the
interactions between the Higgs boson and gauge bosons. They are
not severely constrained by electroweak measurements at the Z pole
or at lower energies~\cite{eboli_concha,dawson}.

The couplings $\d$ and $\db$ were introduced in
Reference~\citen{gounaris}, while $\dg1z$ and $\dkg$ also describe
possible deviations in the couplings of W bosons with photons and Z
bosons~\cite{hagiwara_ww}.  A search for anomalous Higgs production
and decay with non-vanishing values of $\dg1z$ or $\dkg$ is a
complementary study to the analysis of triple-gauge-boson couplings in
the $\ee\ra\W^+\W^-$ process.  The parameter
$\xi^2=(1+\delta_\Zo)^2$ describes a global rescaling of all
Higgs couplings and affects the Higgs production cross section, but
not its branching fractions~\cite{wudka}.

\begin{figure}[htbp]
\begin{center}
    \includegraphics[width=0.8\textwidth]{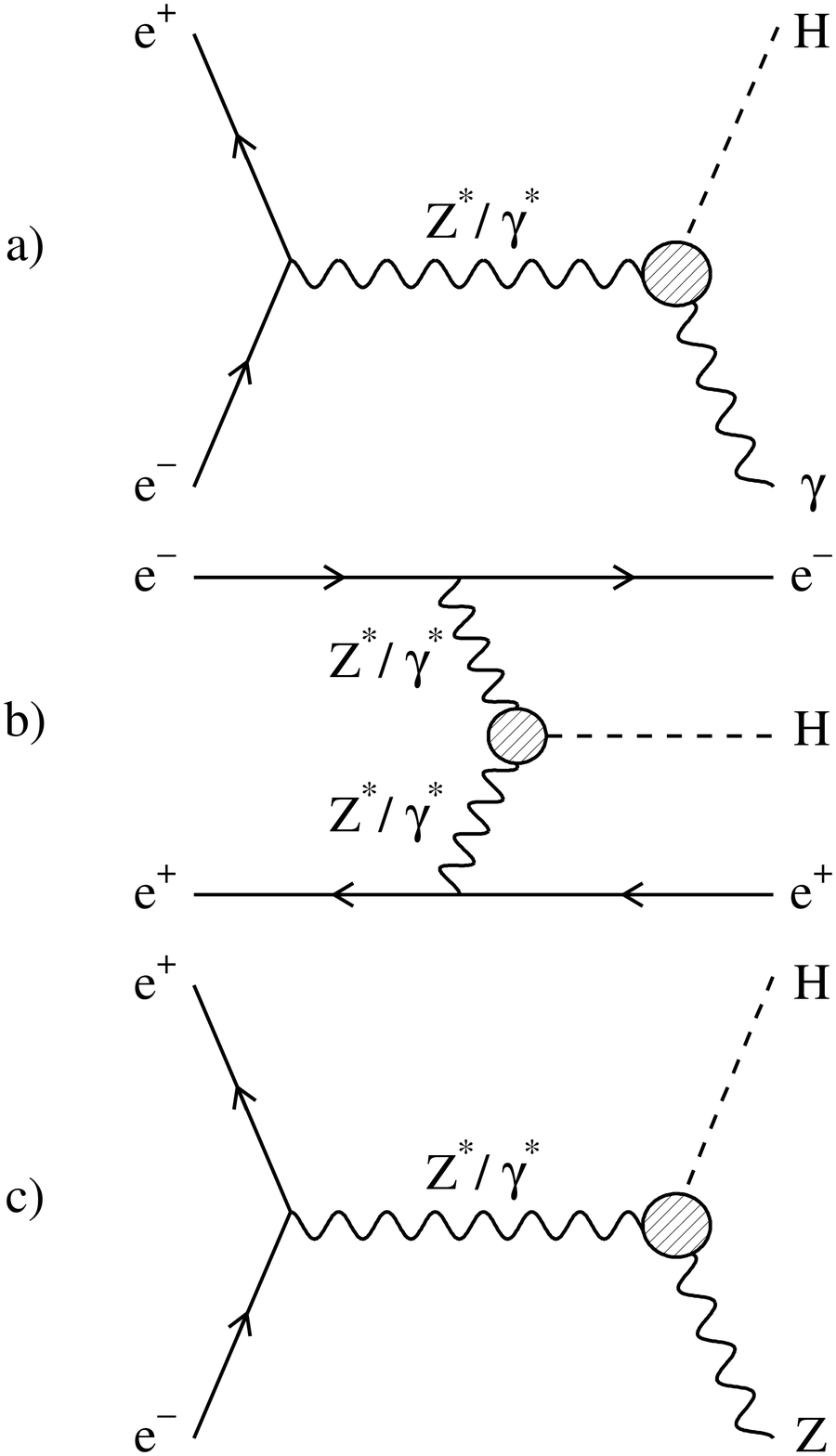}
\end{center}
\caption{
    Relevant production processes in the search for
anomalous couplings in the Higgs sector at LEP: a) $\eehg$, b) $\eehee$ and c) $\eehz$.}
\label{figure:diagrams}
\end{figure}

We search for a Higgs particle produced in the $\eehg$ and $\eehee$
processes shown in Figures~\ref{figure:diagrams}a
and~\ref{figure:diagrams}b. Their rates would be enhanced in presence
of anomalous $\HGG$ and $\HZG$ couplings.  These processes probe Higgs
masses, $\MH$, up to the centre-of-mass energy of the collision, $\rts$.
For $\MH < \rts-\MZ$, this analysis is complemented by the
results from the L3 searches for the $\eehz$
process~\cite{l3sm_paper,l3aura_paper}, which are sensitive to
anomalous $\HZZ$ and $\HZG$ couplings, as shown in Figure
\ref{figure:diagrams}c.

The existence of $\HGG$ and $\HZG$ couplings would lead to large
$\htogg$ and $\htozg$ branching fractions, which at tree level are
zero in the Standard Model. These decay modes have complementary
sensitivities and allow to probe a large part of the parameter
space. In addition, the decay $\htoww$ would also be enhanced in the
presence of anomalous $\HWW$ couplings.

The data used in this analysis were collected with the L3
detector~\cite{L3DET} at LEP at $\rts=189- 209 \GeV$ and correspond to
an integrated luminosity of 602\,$\pb$. Searches for anomalous Higgs
production were previously performed, with data of lower energy and
integrated luminosity, by L3 and other
experiments~\cite{L3_anom,hg_delphi}. Other non-standard Higgs
searches performed at LEP are reported in~\cite{l3aura_paper,LEP_fermio}. The
results reported in this Letter include and supersede those of
Reference~\citen{L3_anom}.

\section {Analysis strategy}

Table \ref{tab:signatures} summarizes the experimental signatures
considered for the study of Higgs anomalous couplings according to the
different production mechanisms and decay channels.

For the $\eehg$ process, the decay channels $\htogg$, $\htozg$ and
$\htoww$ are investigated. Only hadronic decays of Z and W bosons
are considered.

For the $\eehee$ process, only the $\htogg$ decay is studied. The
$\htobb$ decay was considered in the study of the $\eehee$ and $\eehg$
processes for data collected at $\rts=189\GeV$~\cite{L3_anom}. This
decay is dominant for $\MH \lesssim \MZ$, where $\htogg$ is strongly
suppressed and $\htozg$ is kinematically forbidden. At the
centre-of-mass energies considered in this Letter, this region is
efficiently covered by an interpretation of the results of the search
for the $\eehz$ process~\cite{l3sm_paper} and the $\htobb$ decay is not considered here.

No dedicated selection is devised for the $\eehz$ process and the
limits obtained by L3 in the searches for the Standard Model Higgs
boson and for a fermiophobic Higgs boson are interpreted in terms
of anomalous Higgs couplings.

The analysis is performed as a function of $\MH$ in steps of $1
\GeV$. The $\htogg$, $\htozg$ and $\htoww$ decays probe the ranges $70
\GeV < \MH < 190 \GeV$, $95 \GeV < \MH < 190 \GeV$ and $130 \GeV < \MH
< 190 \GeV$, respectively.

After the event selections described below, variables which depend on
$\MH$ are built to discriminate signal and background.  Finally, the
number of events in a mass window around the $\MH$ value under study
is compared with the Standard Model expectation and interpreted in
terms of cross sections and anomalous couplings.

\section {Data and Monte Carlo samples}

Table \ref{tab:energies} lists the centre-of-mass energies and the
corresponding integrated luminosities used in this analysis.
The data at $\rts = 189 \GeV$ are re-analysed for the $\eeggg$ and
$\eeeegg$
channels and results for the full range $\rts = 189 - 209 \GeV$ are
reported here. All other analyses discussed in this Letter refer to
the $\rts = 192 - 209 \GeV$ range, and their results are then combined
with those obtained at $\rts = 189 \GeV$\cite{L3_anom}.

To describe the $\eehg$ process we wrote a Monte Carlo 
generator which assumes a $1+\cos^2\theta_\Ho$ dependence of the
differential cross section as a function of the cosine of the
Higgs production angle, $\theta_\Ho$. It includes effects of
initial-state~\cite{remt} and final-state~\cite{photos} radiation
as well as spin correlations and off-shell contributions in
cascade decays such as $\Ho\ra\Zo\gamma\ra\ffbar\gamma$.

The $\eehee$ process is interpreted as the production of a
narrow-width spin-zero resonance in two-photon collisions, and
modelled with the PC Monte Carlo generator \cite{frank_linde}.

The differential cross section of the process $\eehz$ in the
presence of anomalous couplings is taken from Reference
\citen{hagiwara_hz}. References \citen{partial_widths} and
\citen{romao} are used for the branching fractions and partial
widths of a Higgs boson with anomalous couplings. The interference
between the $\eehz$ process in the Standard Model and in presence of
anomalous couplings~\cite{hagiwara_hz} is taken into account in the
simulation. It is negligible for the $\eehg$ and
$\eehee$ cases.  

Signal events are generated for $70 \GeV < \MH < 190 \GeV$, in
steps of $20 \GeV$. More than 5000 signal events are generated for
each value of $\MH$ and for each process under study. For
intermediate values of the Higgs mass, the signal efficiency is
interpolated between the generated values.

Standard Model processes are modelled with the following Monte
Carlo generators: GGG~\cite{ggg} for $\ee\ra\gamma\gamma(\gamma)$,
KK2f \cite{kk2f} for $\ee\ra {\rm q}{\bar{\rm q}}(\gamma)$, PYTHIA
\cite{pythia} for $\ee\ra \Zo\Zo$ and $\ee\ra \Zo\ee$,
KORALW \cite{koralw_1} for $\ee\ra\W^+\W^-(\gamma)$ and EXCALIBUR
\cite{excalibur} for $\ee\ra {\rm We}\nu$ and other four-fermion
final states.

The L3 detector response is simulated using the GEANT
program~\cite{my_geant} which takes into account effects of energy
loss, multiple scattering and showering in the detector.
Time-dependent detector inefficiencies, as monitored during the data-taking
period, are included in the simulations.

\section{Event selection}

All analyses presented in this Letter rely on photon
identification. Photon candidates are defined as clusters in the
electromagnetic calorimeter with a shower profile consistent with that
of a photon and no associated track in the tracking chamber. To
reduce contributions from initial-state and final-state radiation,
photon candidates must satisfy $E_\gamma > 5 \GeV$ and
$\mid\cos\theta_\gamma\mid < 0.97$, where $E_\gamma$ is the photon
energy and $\theta_\gamma$ its polar angle.

Events with hadronic decays of the Z and W bosons in the $\htozg$
and $\htoww$ channels are pre-selected requiring high particle
multiplicity and a visible energy, ${E}_{vis}$, satisfying $0.8 <
{E}_{vis}/\rts < 1.2$.

\subsection{The \boldmath{$\eeggg$} analysis }

\begin{figure}[htbp]
\begin{center}

\begin{tabular}{cc}
   \includegraphics[width=0.5\textwidth]{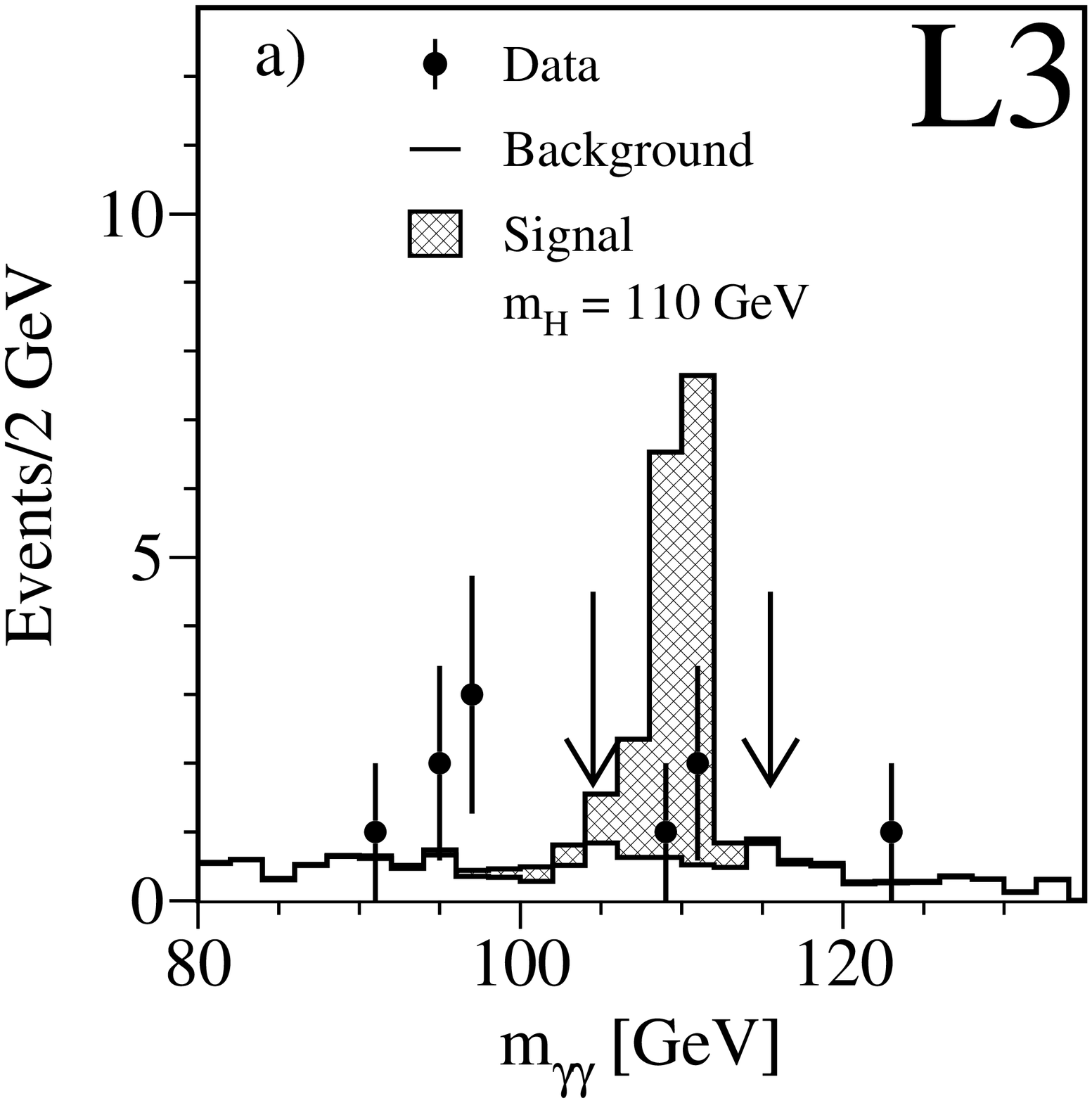} &
    \includegraphics[width=0.5\textwidth]{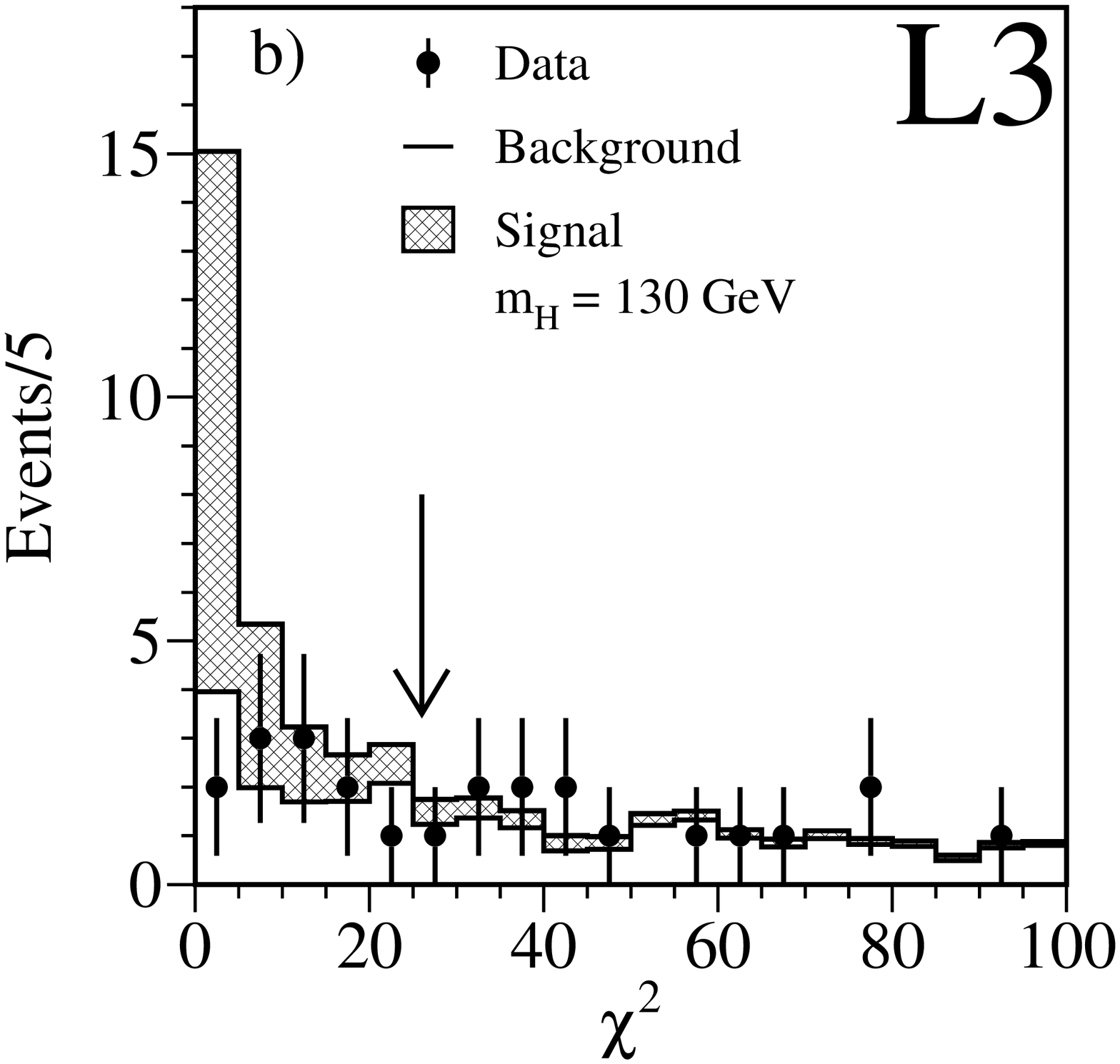} \\
    \includegraphics[width=0.5\textwidth]{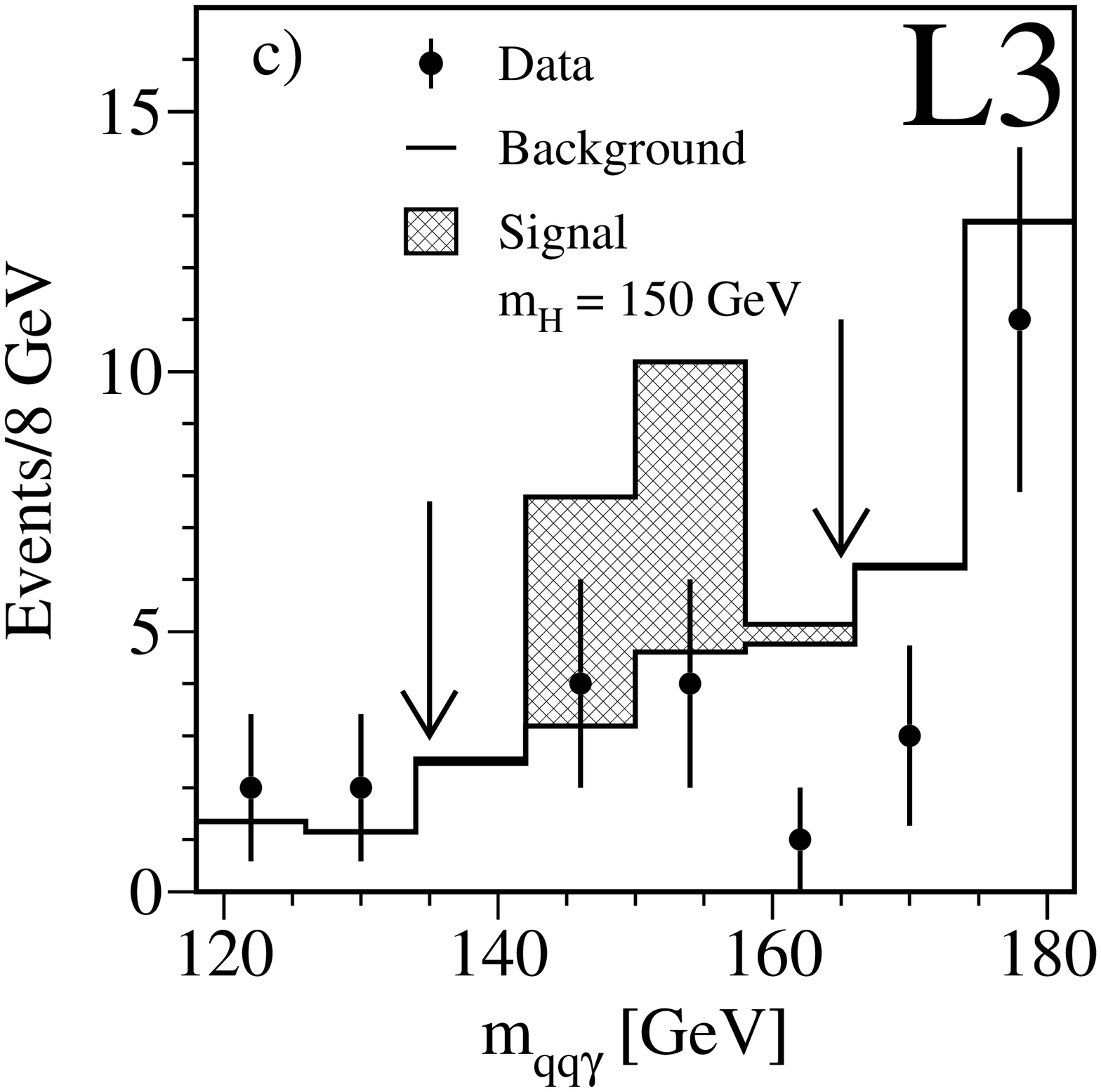} &
    \includegraphics[width=0.5\textwidth]{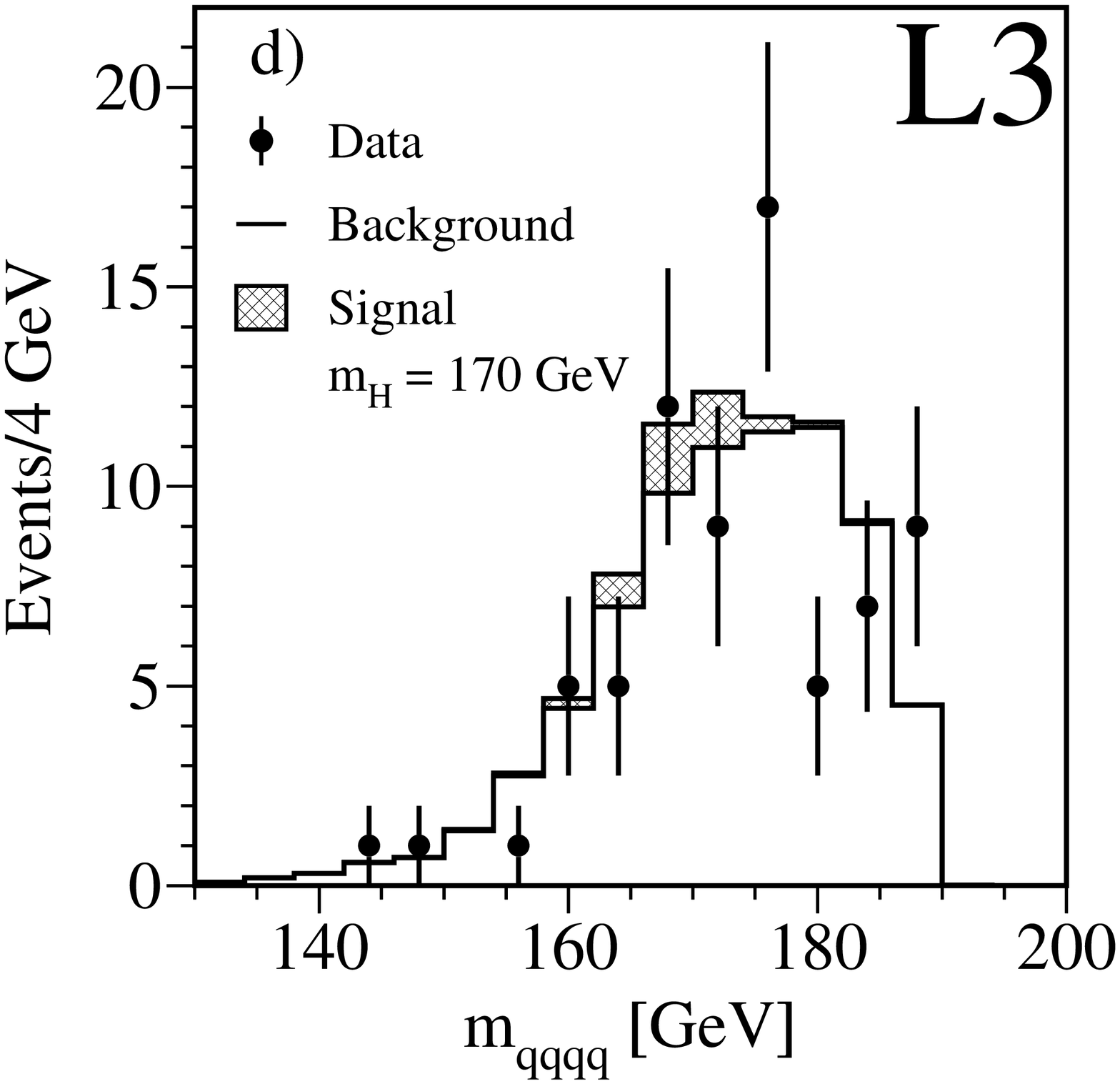}      \\
\end{tabular}
\end{center}
\caption{Distributions of the final discriminant variables for
a) the $\eetoggg$ channel: the mass, $m_{\gamma\gamma}$, of the two-photon system;
b) the $\eetoeegg$ channel: the $\chi^2$ of the constrained fit;
c) the $\eetozgg$  channel: the mass, $m_{\rm qq\gamma}$, of the system of the two-jets and
a photon
and d) the $\eetowwg$  channel: the mass, $m_{\rm qqqq}$, of the hadronic system. The
points represent the data, the open histograms the background and the
hatched histograms the Higgs signal with an arbitrary cross section of 0.1 pb.
The Higgs mass hypotheses indicated
in the figures are considered. The arrows indicate the
values of the cuts.
}
\label{fig:exp_plots}
\end{figure}

Events from the $\eeggg$ process are selected by requiring three
photon candidates in the central region of the detector,
$|\cos\theta_\gamma| < 0.8$, with a total electromagnetic energy
larger than $\rts/2$. Out of the three possible two-photon
combinations, the one with a mass, $m_{\gamma\gamma}$, closest to
the $\MH$ hypothesis under investigation is retained. As an
example, Figure \ref{fig:exp_plots}a presents the distribution of
$m_{\gamma\gamma}$ for $\MH=110 \GeV$.  The event is accepted as a
Higgs candidate if $\mid m_{\gamma\gamma}-\MH\mid < 0.05 \,\MH$.

The numbers of events observed and expected in the full data
sample at $\rts=189-209 \GeV$ are shown in Table \ref{tab:events}
for several $\MH$ hypotheses. The contamination from processes
other than $\ee\ra\gamma\gamma(\gamma)$, as estimated from Monte
Carlo simulations, is found to be negligible.  The signal
selection efficiency
is in the range $25\%-30\%$,
depending on $\MH$ and $\rts$.

\subsection{The \boldmath{$\eeeegg$} analysis}

In the process $\eehee$, the final state $\em$ and $\ep$ tend to escape
detection at low polar angles, originating events with missing
longitudinal momentum and missing mass. The selection requires two
photon candidates from the $\htogg$ decay in the central region of
the detector.  A kinematic fit is performed assuming the missing
momentum to point in the beam pipe and the visible mass of the
event to be consistent, within the experimental resolution, with
the $\MH$ hypothesis under investigation. The distribution of the
$\chi^2$ of the fit is shown in Figure \ref{fig:exp_plots}b for
$\MH = 130 \GeV$. Events are accepted as Higgs candidates if
$\chi^2 < 50 - 0.2 \GeV^{-1}\times\MH$.
The dependence of the cut on $\MH$ reflects the decrease of the background
contribution for increasing values of $\MH$.

The numbers of events observed and expected in the full data
sample at $\rts=189-209 \GeV$ are shown in Table \ref{tab:events}
for several $\MH$ hypotheses. The background comes from
$\ee\ra\gamma\gamma(\gamma)$ events.  The signal selection efficiency
varies from $20\%$ to $30\%$, with a smooth
dependence on $\MH$ and $\rts$.

\subsection{The \boldmath{$\eezgg$} analysis }

Pre-selected hadronic events with two isolated high energy photons
are considered for the $\eezgg$ analysis. Events are retained which have
a recoiling mass, $m_{rec}$, calculated from the four-momenta of the
two photons, compatible with $m_{\rm Z}$: $80 \GeV < m_{rec} < 110 \GeV$. 
The hadronic system
is clustered into two jets with the DURHAM~\cite{durham}
algorithm and a kinematic fit, in which the jet angles are fixed and
the jet energies can vary, is performed to improve the resolution
on the reconstructed Z-boson mass. Of the two possible combinations of
two jets and a photon, the one is retained with mass, $m_{\rm
qq\gamma}$, closer to the $\MH$ hypothesis under investigation.
The distribution of $m_{\rm qq\gamma}$ is shown in Figure
\ref{fig:exp_plots}c for $\MH = 150 \GeV$.  An event is considered
as a Higgs candidate if $|m_{\rm qq\gamma} - \MH| <15 \GeV$.

The numbers of events observed and expected in the data sample at
$\rts=192-209 \GeV$ are shown in Table \ref{tab:events} for several
$\MH$ hypotheses. The signal selection efficiency is
around $22\%$.  The background is dominated by resonant $\ee\ra
\Zo\gamma\gamma$ production (70\%) with contributions from the $\ee\ra {\rm
q}{\bar{\rm q}}(\gamma)$ process and four-fermion final states.

\subsection{The \boldmath{$\eewwg$} analysis}

The energy of the photon in the $\eewwg$ process depends on $\MH$
as $E^{rec}_\gamma (\MH) =(s - \MH^2)/2\rts$. Pre-selected hadronic events are
retained if they have a photon with energy compatible with the $\MH$ hypothesis
under investigation,
$E^{rec}_\gamma (\MH+20 \GeV) < E_\gamma < E^{rec}_\gamma (\MH-20 \GeV)$.
If multiple photon candidates are observed, the
photon is retained which has an energy closest to $E^{rec}_\gamma$($\MH$).
The rest of the event is clustered into four jets by means of the DURHAM
algorithm.

A kinematic fit, in which the jet angles are fixed and the jet
energies can vary, is performed to improve the resolution on the
reconstructed W-boson mass. For $\MH > 2\MW$ both W bosons are on-shell and
the constraint that both invariant jet-jet masses be compatible with
$\MW$ is included in the fit.  For $\MH < 2\MW$ one of the W bosons is
off-shell and only one of the invariant jet-jet masses is required to
be compatible with $\MW$. The fit is repeated for all possible jet
pairings and the pairing is chosen for which the $\chi^2$ of the fit
is minimal.  An event is considered as a Higgs candidate if $\chi^2 <
6.0$ for the hypothesis $\MH < 2\MW$ or $\chi^2 < 15.0$ for $\MH >
2\MW$.  The invariant mass of the four-jet system, $m_{\rm qqqq}$,
estimates $\MH$. Its distribution is presented in Figure
\ref{fig:exp_plots}d for $\MH = 170 \GeV$.

The numbers of events observed and expected in the data sample at
$\rts=192-209 \GeV$ are shown in Table \ref{tab:events} for several
$\MH$ hypotheses.  The signal selection efficiency is
around $25\%$, for $150 \GeV < \MH< 170 \GeV$, decreasing to
about 20\% for masses out of this range. A small dependence on
$\rts$ is observed.  The background is dominated by the processes
$\ee\ra {\rm q}{\bar{\rm q}}(\gamma)$ and
$\ee\ra\W^+\W^-(\gamma)$, which is above $65\%$ for $\MH >150 \GeV$.

\section{Cross sections limits}

The results of all the analyses agree with the Standard Model
predictions and show no evidence for a Higgs boson with anomalous
couplings in the $\MH$ mass range under study.  Upper limits on
the product of the production cross sections and the corresponding
decay branching fractions are derived~\cite{lephwg} at the 95\%
confidence level (CL).
The cross section of the $\eehee$ process is proportional to the
partial Higgs width into photons, $\Ghgg$, and limits
are quoted on $\Ghgg\times\Brhgg$.

In order to combine data sets at different $\rts$ values, a dependence
of the type $\sigma^{AC}(\rts) = \zeta \,\sigma^{SM}(\rts)$ is assumed
for the cross section of anomalous Higgs production,
$\sigma^{AC}$. The  $\eehg$ production cross section in the Standard
Model, $\sigma^{SM}$, accounts for the dominant dependence on $\rts$
while $\zeta$ is a parameter which does not depend on $\rts$. Limits
on $\zeta$ are derived and interpreted as cross section limits at the
luminosity-averaged centre-of-mass energy $<\rts> = 197.8 \GeV$.

\begin{figure}[htbp]
\begin{center}
\begin{tabular}{cc}
    \includegraphics[width=0.5\textwidth]{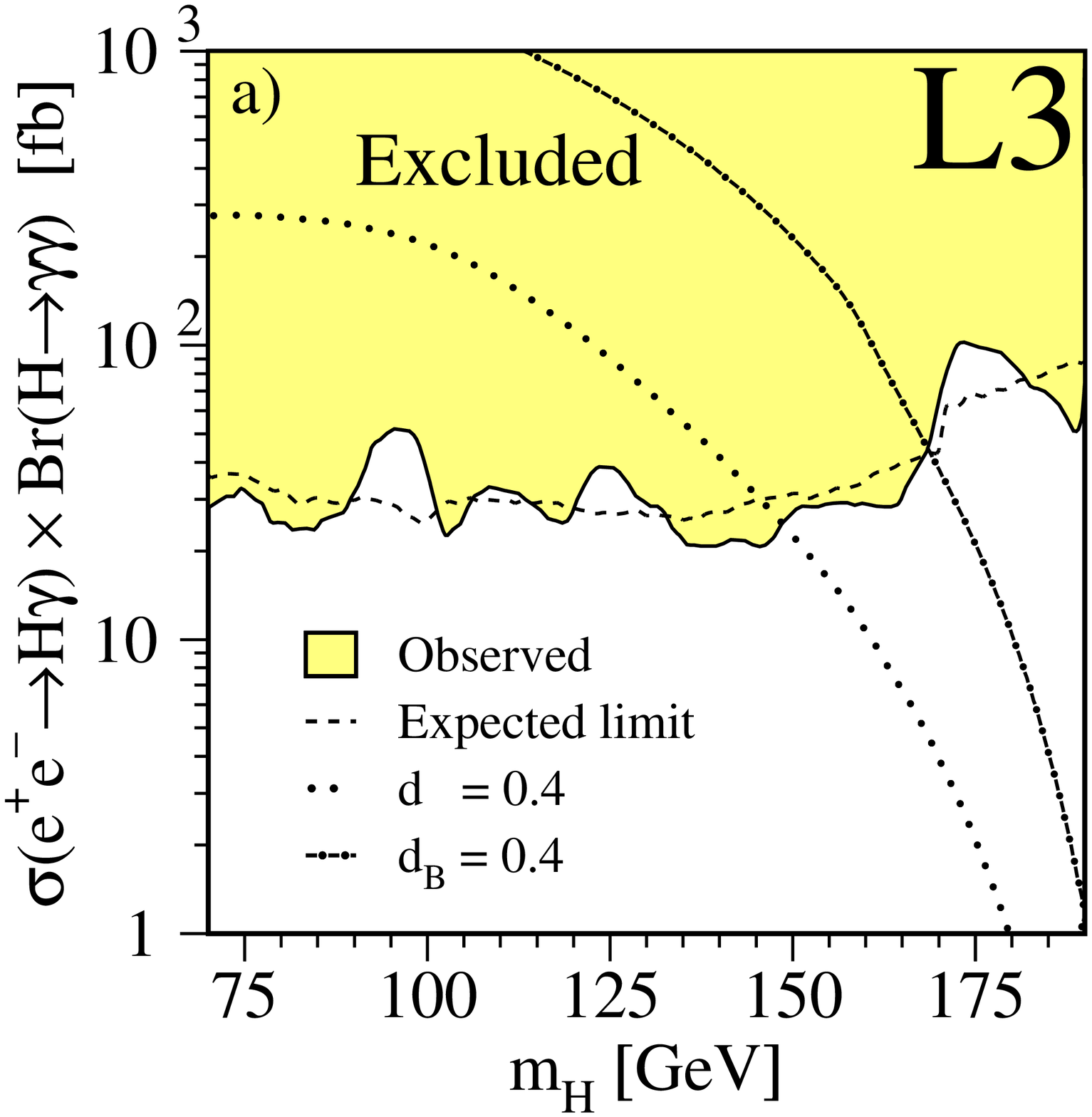} &
    \includegraphics[width=0.5\textwidth]{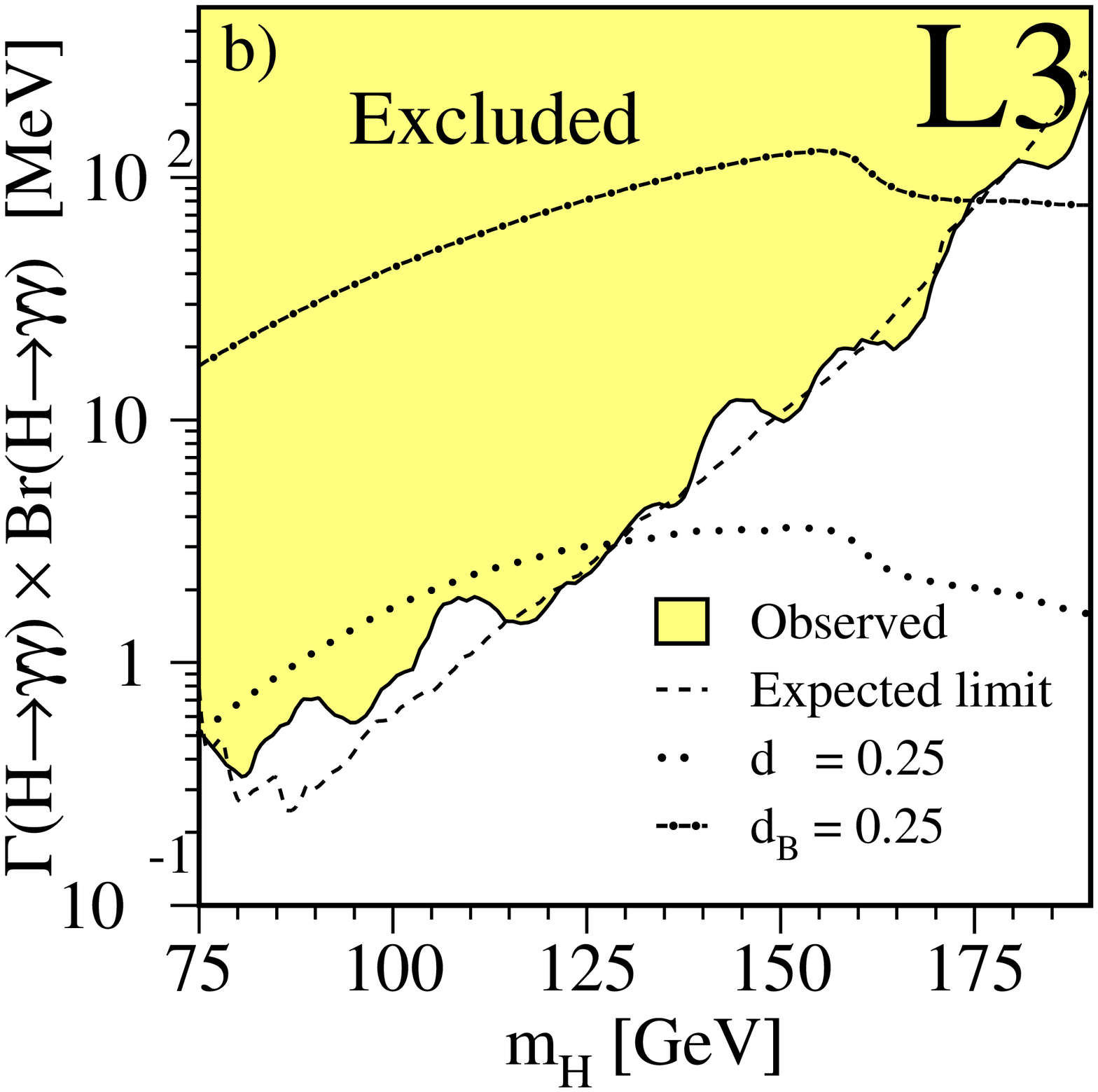}  \\
    \includegraphics[width=0.5\textwidth]{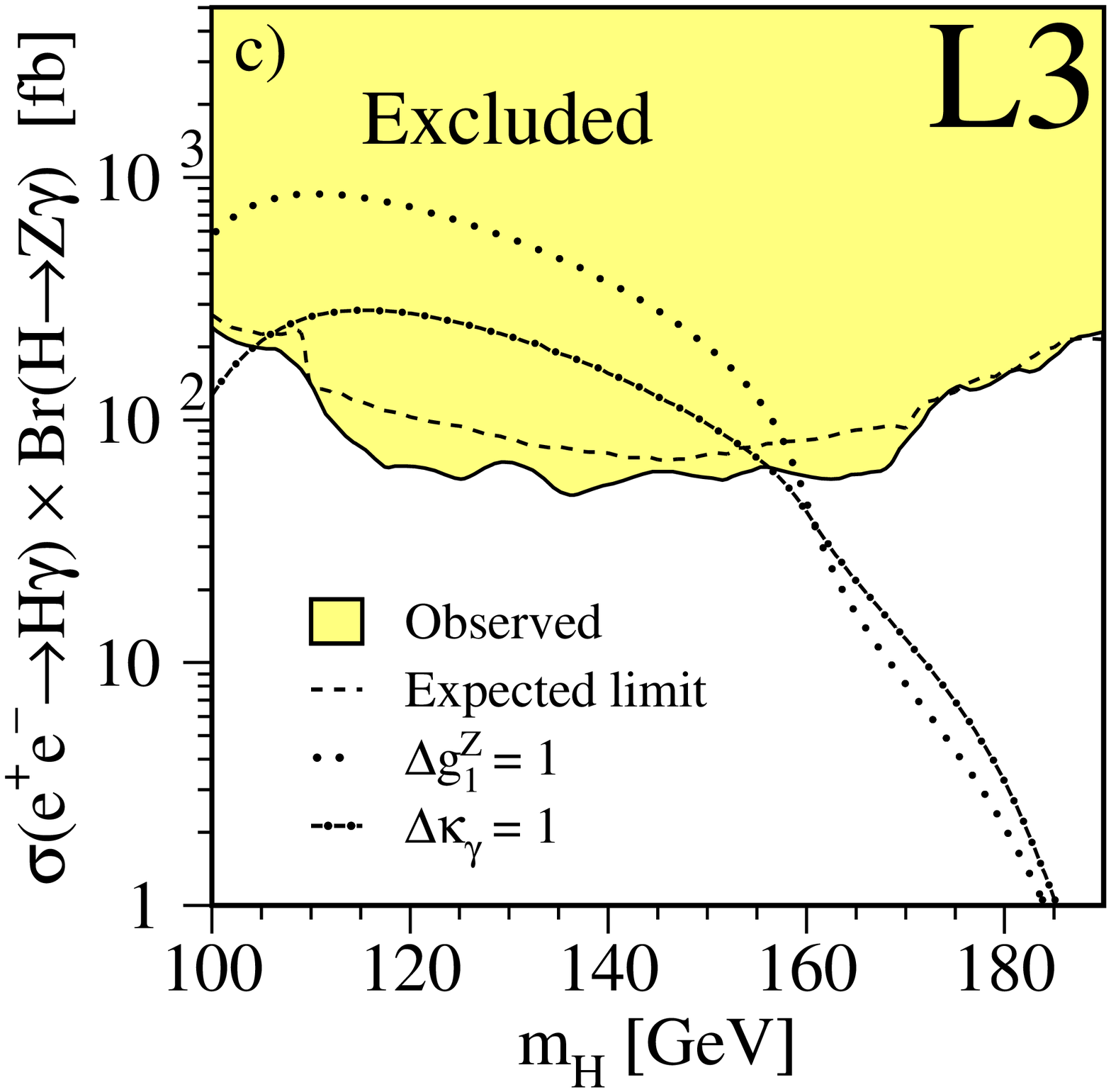} &
    \includegraphics[width=0.5\textwidth]{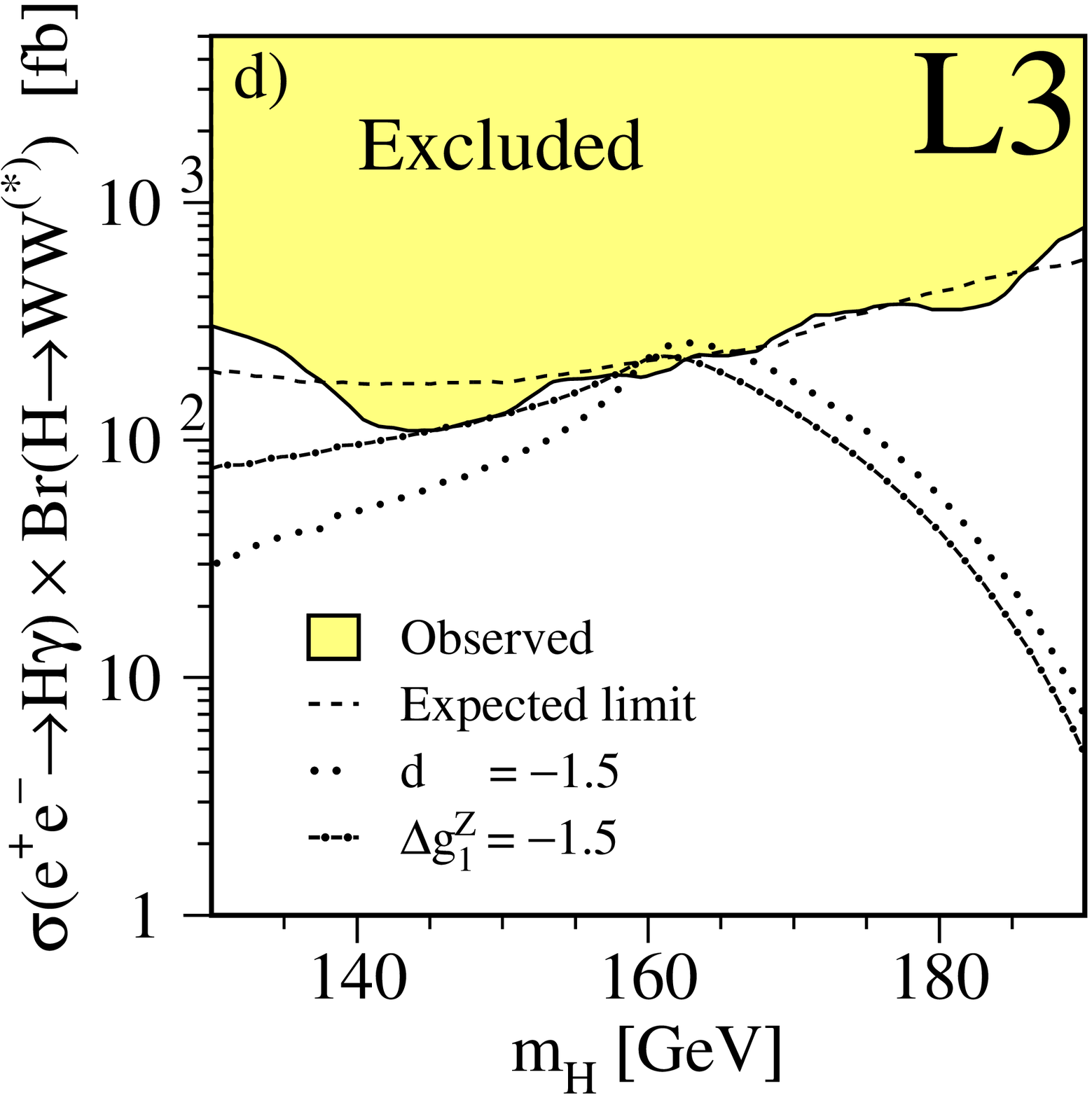} \\
\end{tabular}
\end{center}
\caption{Upper limits at 95\% CL as a function of the Higgs mass
on:
a) $\sigma(\eehg)\times\Brhgg$;
b) $\Ghgg\times\Brhgg$;
c) $\sigma(\eehg)\times\Brhzg$;
d) $\sigma(\eehg)\times\Brhww$.
The dashed line indicates the expected limit in the absence of a
signal. 
Predictions for non-zero values of the anomalous
couplings are also shown.}
\label{fig:xs_limits}
\end{figure}

The cross section limits for the investigated processes are given
in Figure~\ref{fig:xs_limits} together with the expectations for
non-zero values of the anomalous couplings.

\section{Limits on anomalous couplings}

\subsection{Results from \boldmath{$\eehz$} with  \boldmath{$\htoff$} or \boldmath{$\htogg$}}

The process $\eehz$, with $\htoff$, studied in
Reference~\citen{l3sm_paper}, is sensitive to anomalous HZZ and
HZ$\gamma$ couplings in the Higgs production vertex. In addition,
the process $\eehz$ with $\htogg$, object of the search for a
fermiophobic Higgs~\cite{l3aura_paper}, is sensitive to the $\HGG$
coupling in the decay vertex.

\begin{figure}[htbp]
\begin{center}
    \includegraphics[width=12cm]{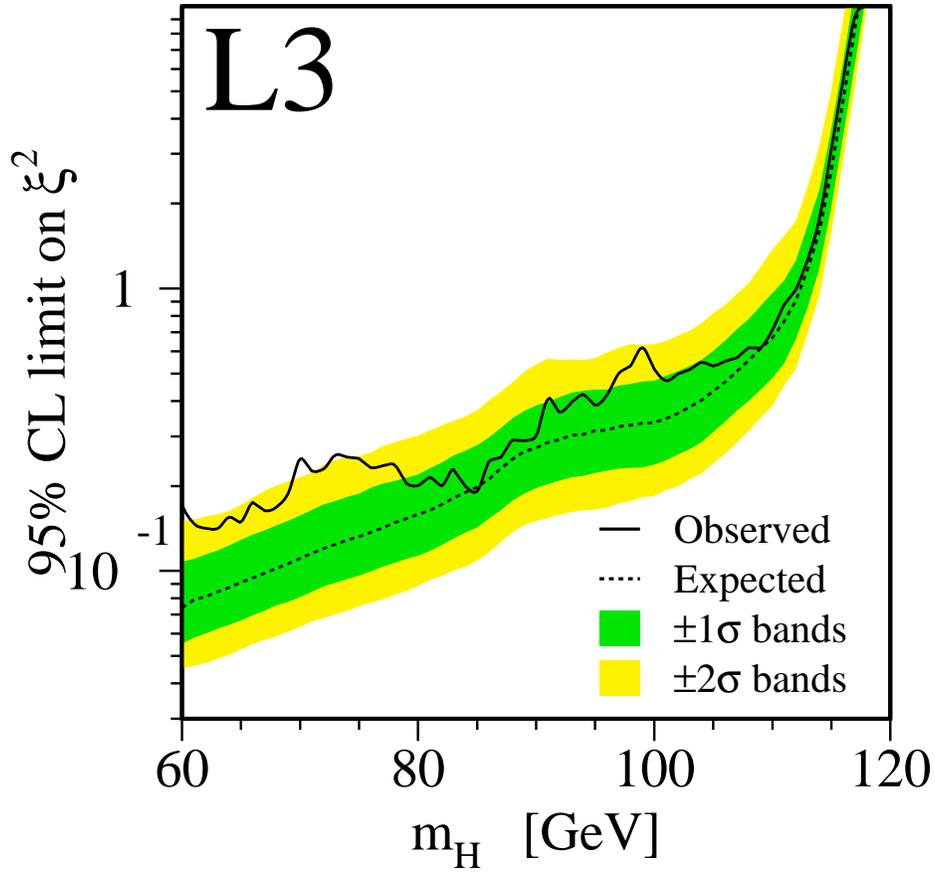}
\end{center}
\caption{The 95\% CL upper bound on the
anomalous coupling $\xi^2$ as
a function of the Higgs mass, as obtained from the results of
the search for the Standard Model Higgs
boson\protect\cite{{l3sm_paper}}. 
The dashed line indicates the expected limit in the absence of a
signal. The dark and light shaded bands around the
expected line correspond to the 68.3\% and 95.4\% probability bands,
denoted by $1\sigma$ and $2\sigma$ respectively.
}
\label{fig:xi_limit}
\end{figure}

Limits on the coupling $\xi^2$ are derived from the results of our
search for the Standard Model Higgs boson~\cite{l3sm_paper}. They
are obtained by interpreting $\xi^2$ as a scale factor of the
Higgs production cross section and are shown in Figure
\ref{fig:xi_limit}. They include the systematic uncertainties on
the search for the Standard Model Higgs boson~\cite{l3sm_paper}.

\begin{figure}[htbp]
\begin{center}
\begin{tabular}{cc}
    \includegraphics[width=0.5\textwidth]{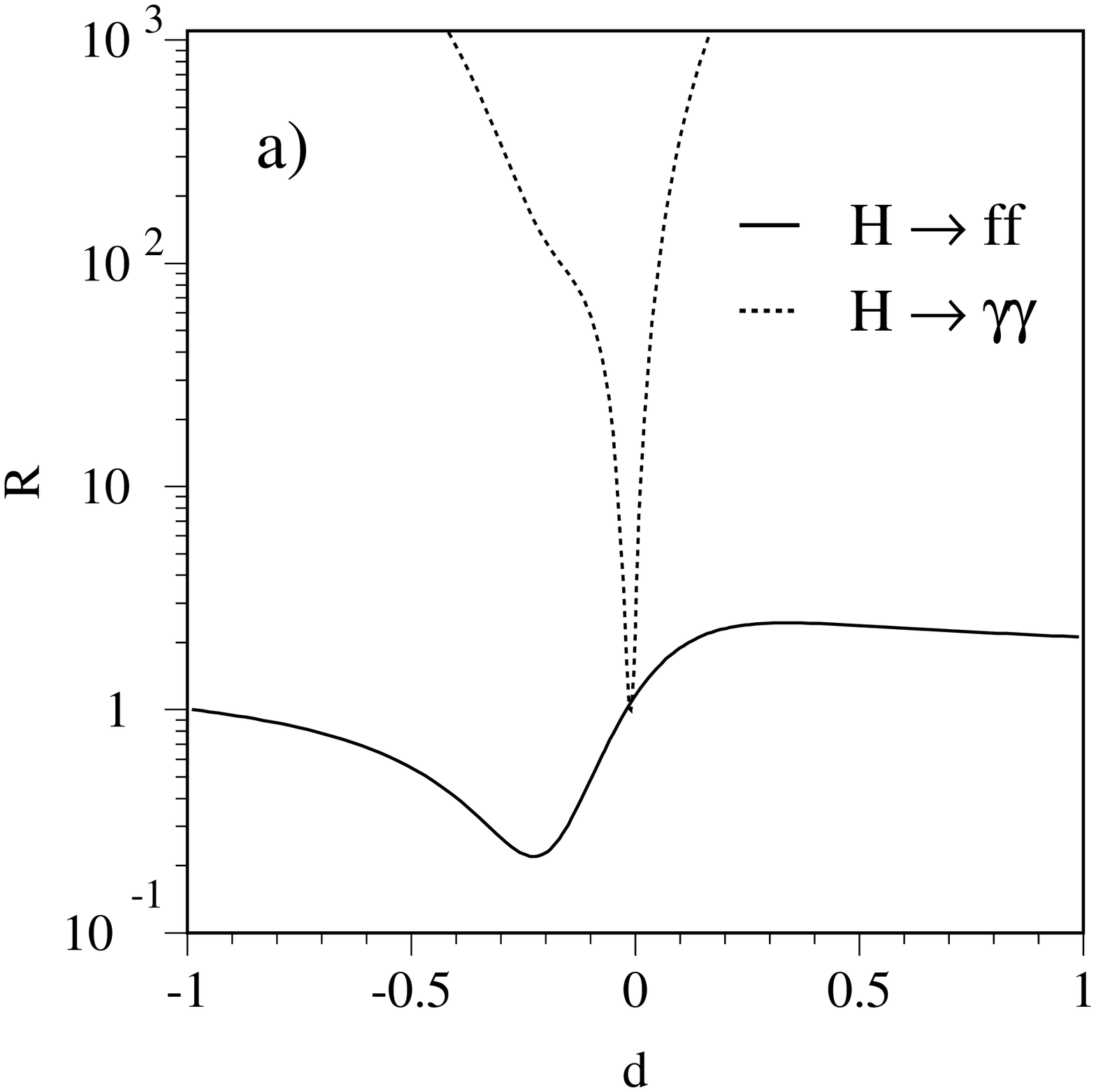} &
    \includegraphics[width=0.5\textwidth]{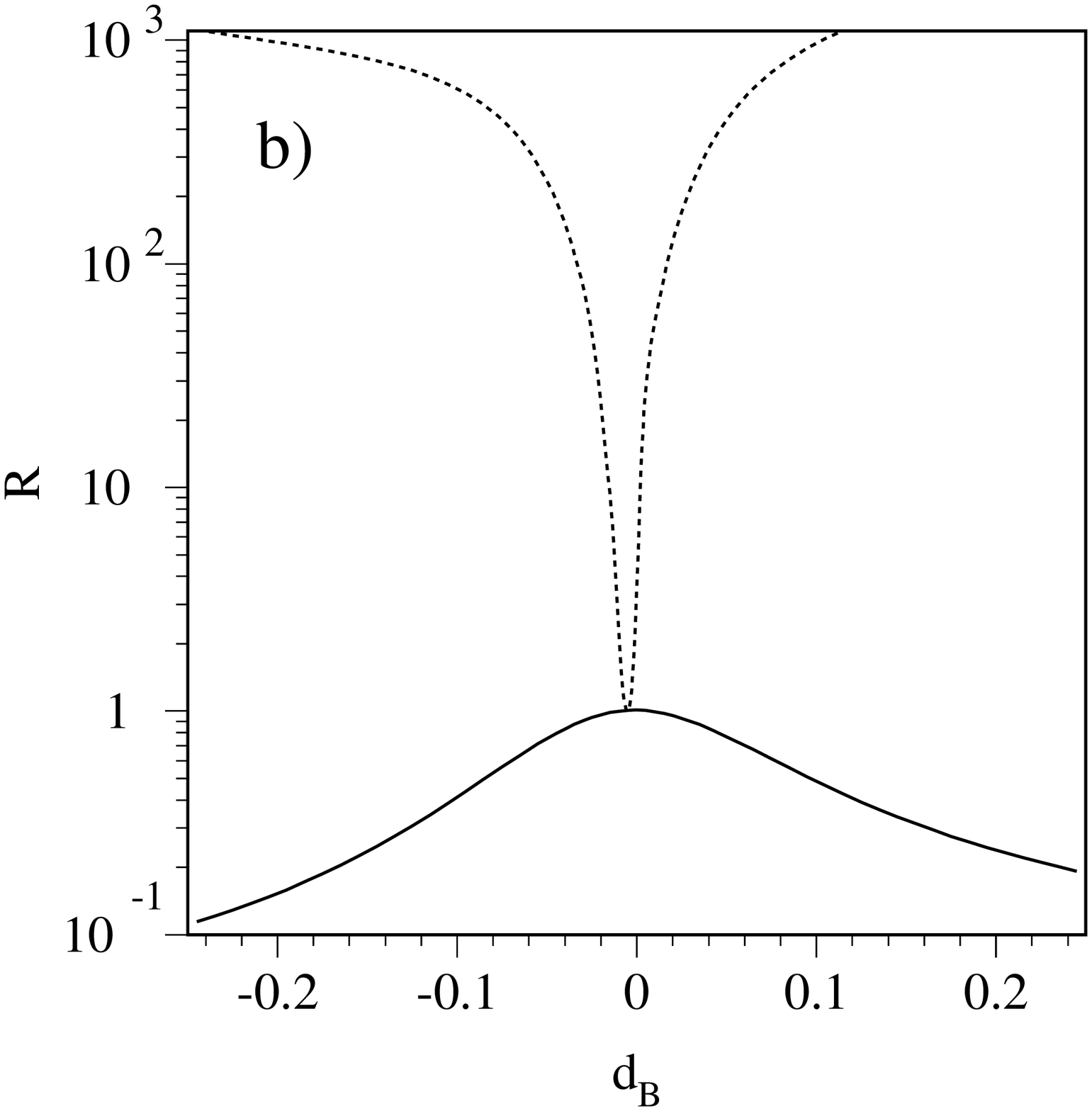}  \\
    \includegraphics[width=0.5\textwidth]{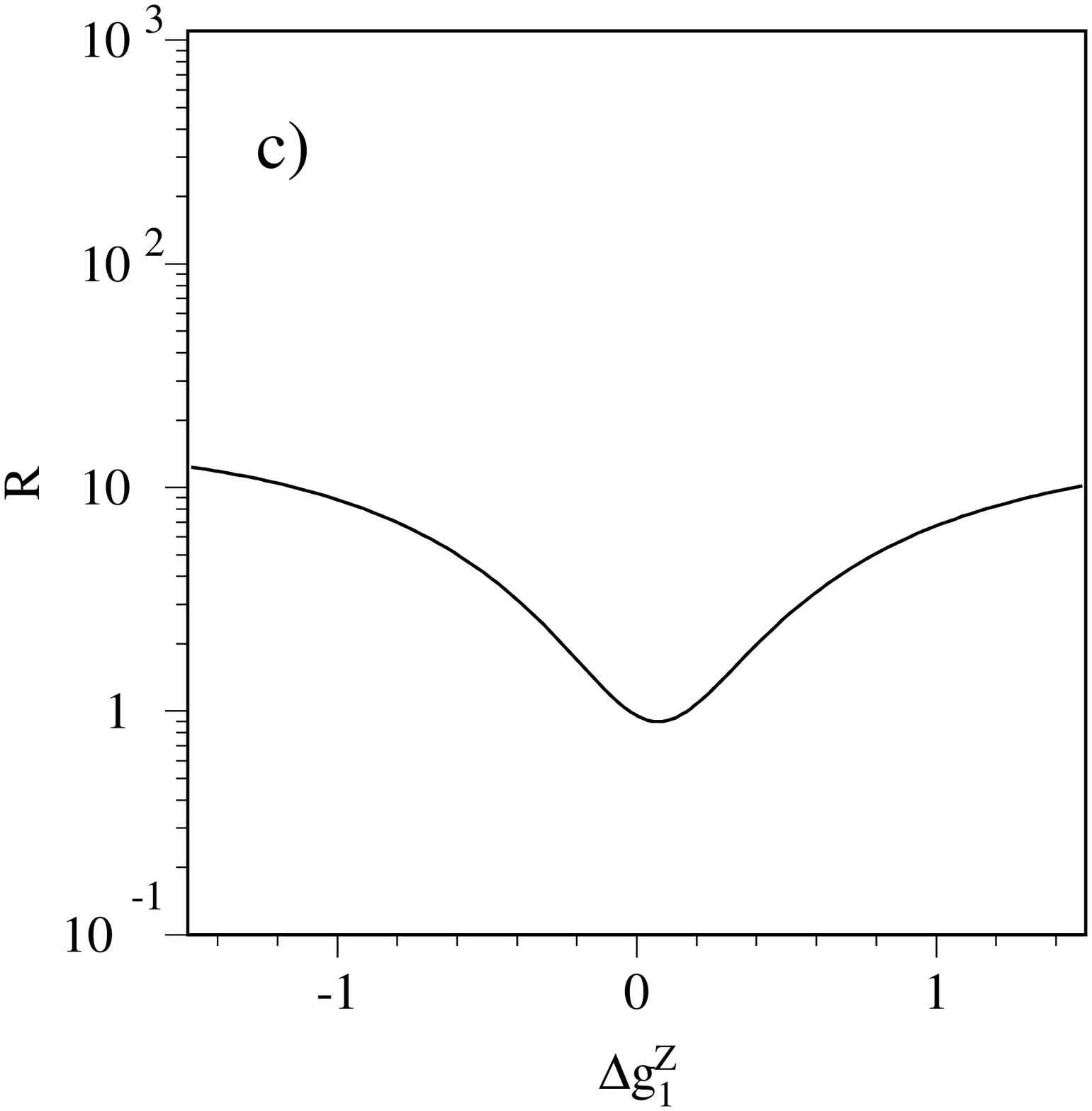} &
    \includegraphics[width=0.5\textwidth]{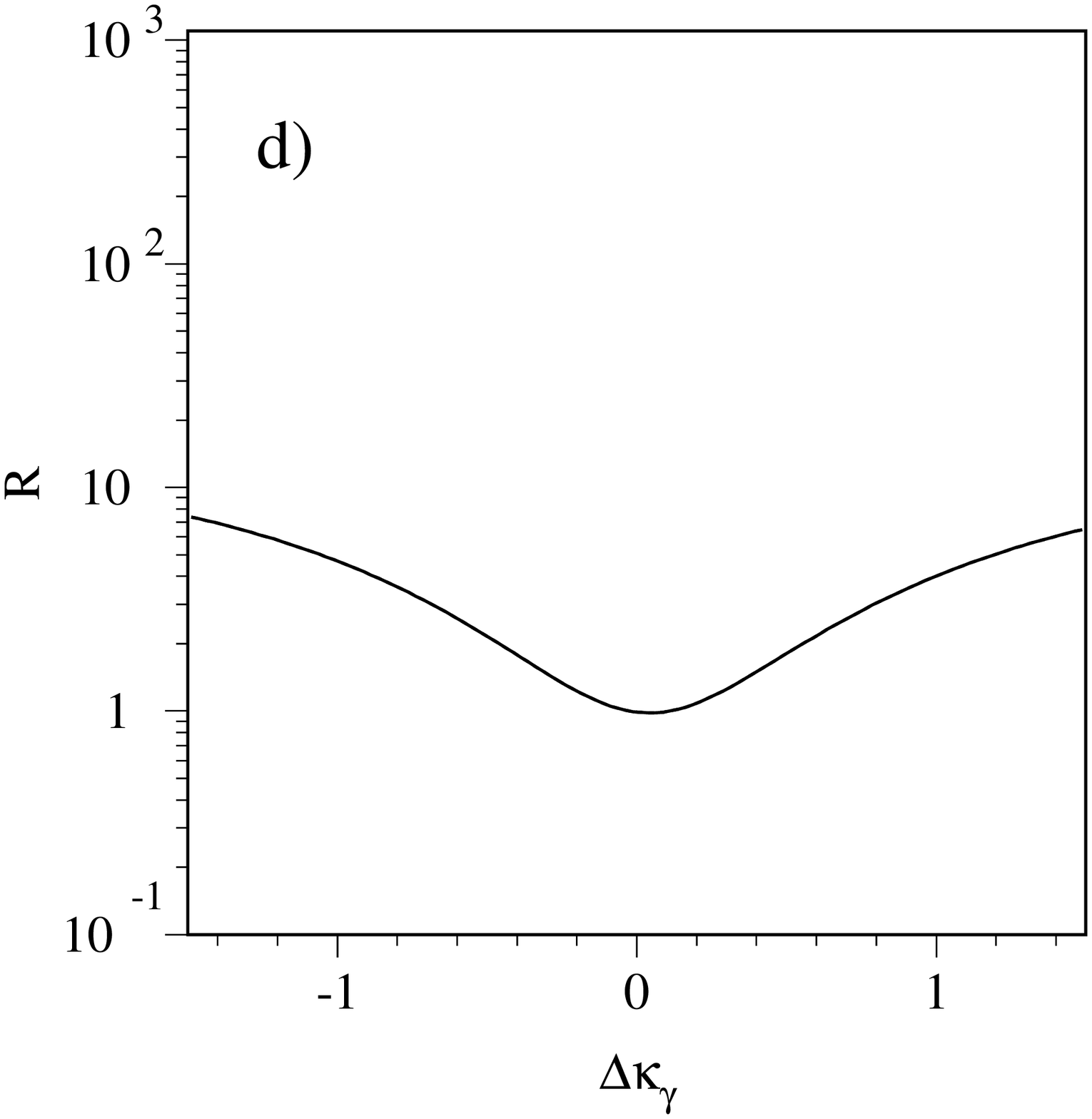} \\
\end{tabular}
\end{center}
\caption{
The theoretical predictions for the
ratios $R = (\sigma^{AC}\times\Br^{AC})/(\sigma^{SM}\times\Br^{SM})$ for the $\eehz$ channel
for the couplings a) $\d$, b) $\db$, c) $\dg1z$ and d) $\dkg$.
The solid line corresponds to the decay $\htoff$ and the dashed line to $\htogg$.
The predictions refer to $\MH=100\GeV$.
The ratios for the two decay modes coincide for $\dg1z$ and $\dkg$.}
\label{fig:ratios}
\end{figure}

The limits on the couplings $\d$, $\db$, $\dg1z$ and $\dkg$ are
extracted from the numbers of observed events, expected background and signal events
reported in References~\citen{l3sm_paper} and~\citen{l3aura_paper}.
These limits are
driven by the size of the deviations of the product $\sigma^{AC}\times\Br^{AC}$
with respect to $\sigma^{SM}\times\Br^{SM}$, where $\Br^{AC}$ and $\Br^{SM}$ denote the
Higgs branching ratios in the 
presence of anomalous couplings and in the Standard Model respectively.
The ratios $R = (\sigma^{AC}\times\Br^{AC})/(\sigma^{SM}\times\Br^{SM})$
are shown in Figure \ref{fig:ratios} for $\htoff$ and $\htogg$, for $\MH=100\GeV$.

The $\htoff$ and $\htogg$ channels have different behaviours with
respect to the parameters $\d$ and $\db$, as these describe the
$\HGG$ coupling. The parameters $\dg1z$ and $\dkg$ describe the
$\HZG$ and $\HZZ$ couplings and hence affect only the Higgs
production vertex in the $\eehz$ process. They give similar
deviations for both the $\htoff$ and $\htogg$ channels.

\subsection{One-dimensional limits}

\begin{figure}[htbp]
\begin{center}
    \includegraphics[width=0.8\textwidth]{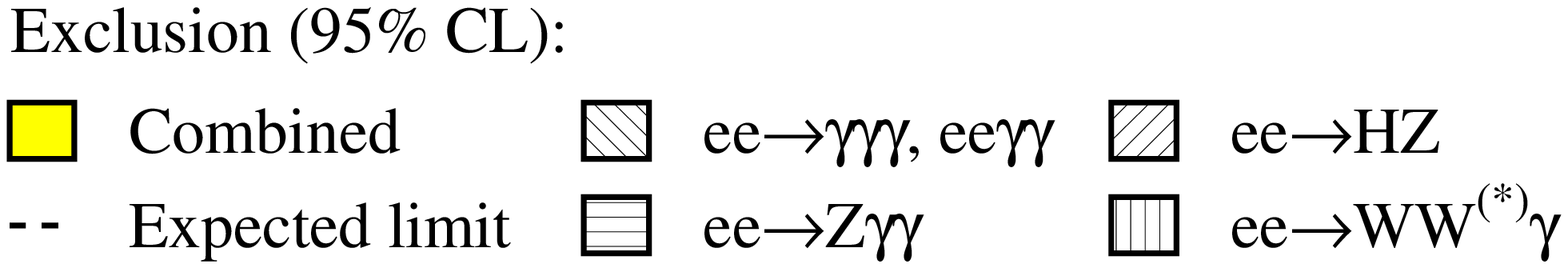}
\vspace*{-0.5cm}
\begin{tabular}{cc}
    \includegraphics[width=0.5\textwidth]{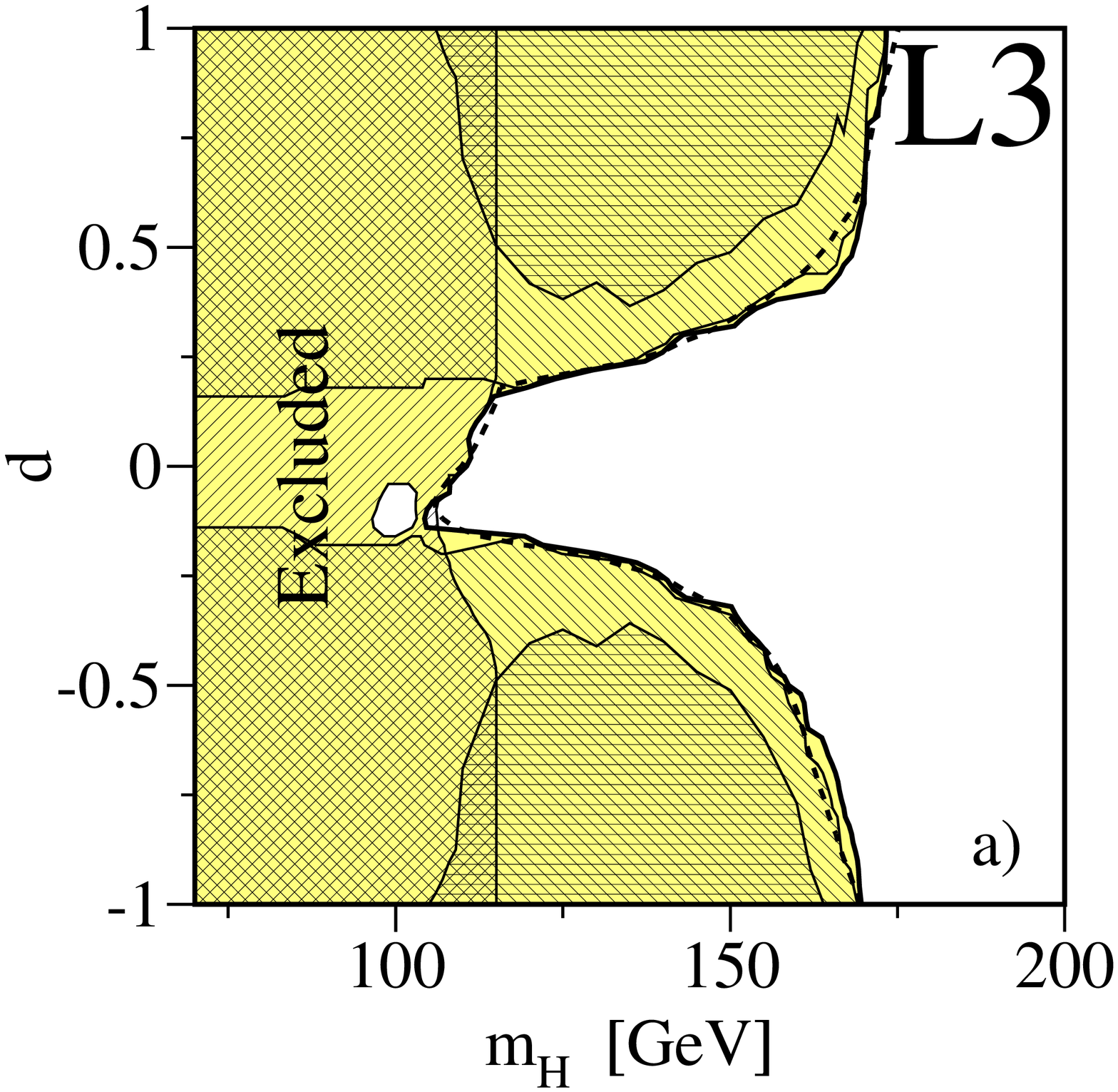} &
   \includegraphics[width=0.5\textwidth] {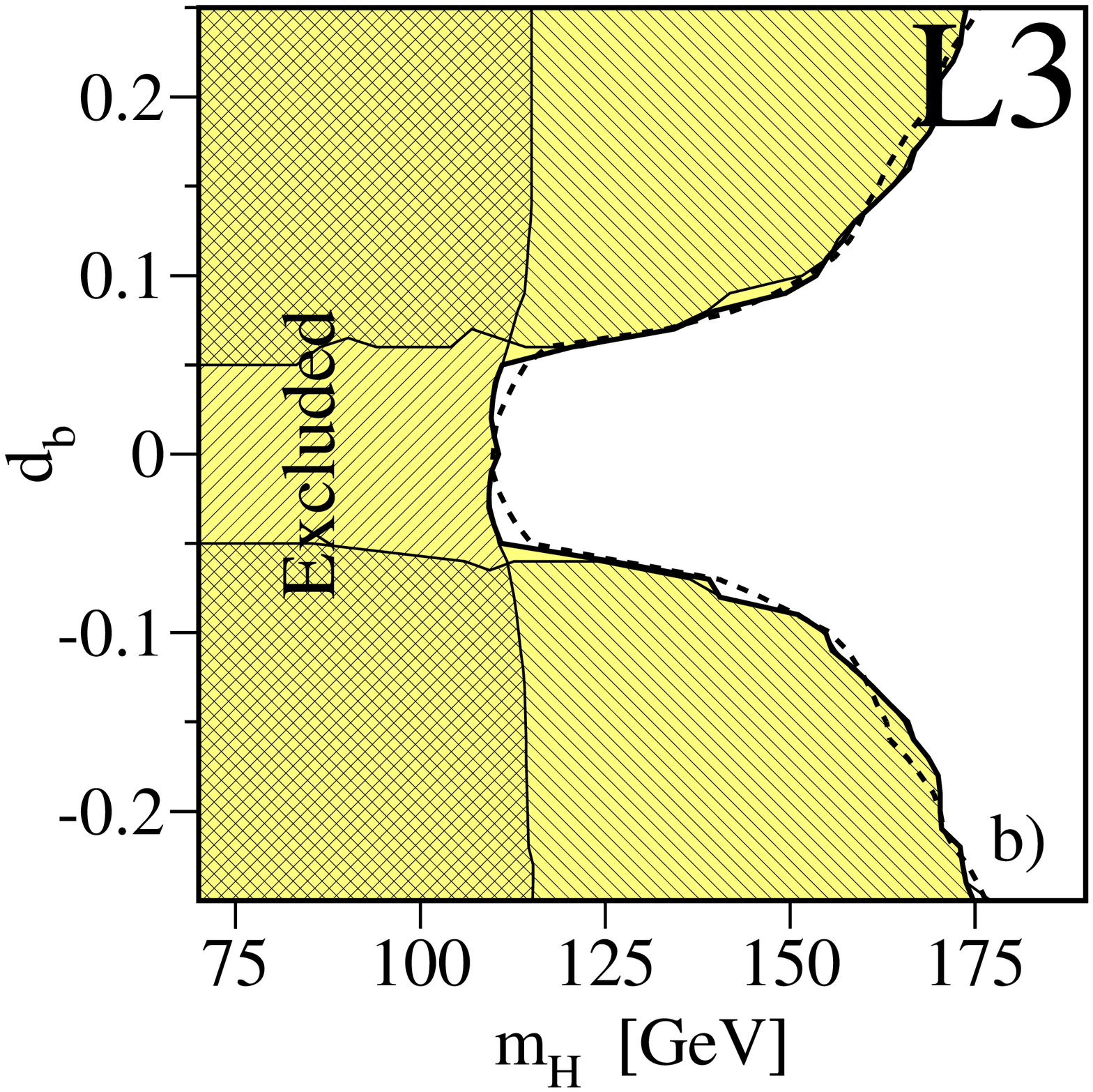} \\
    \includegraphics[width=0.5\textwidth]{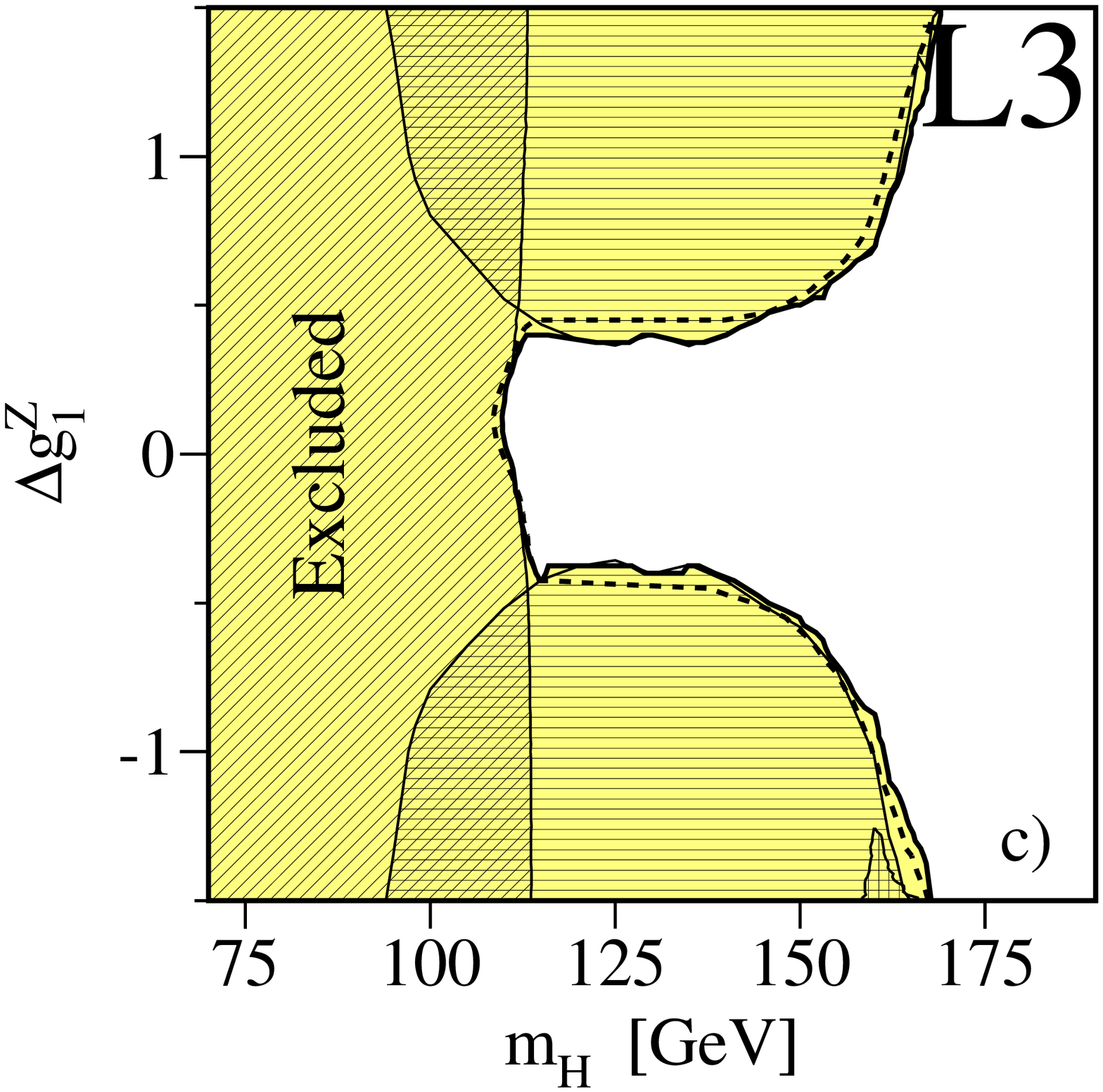} &
   \includegraphics[width=0.5\textwidth] {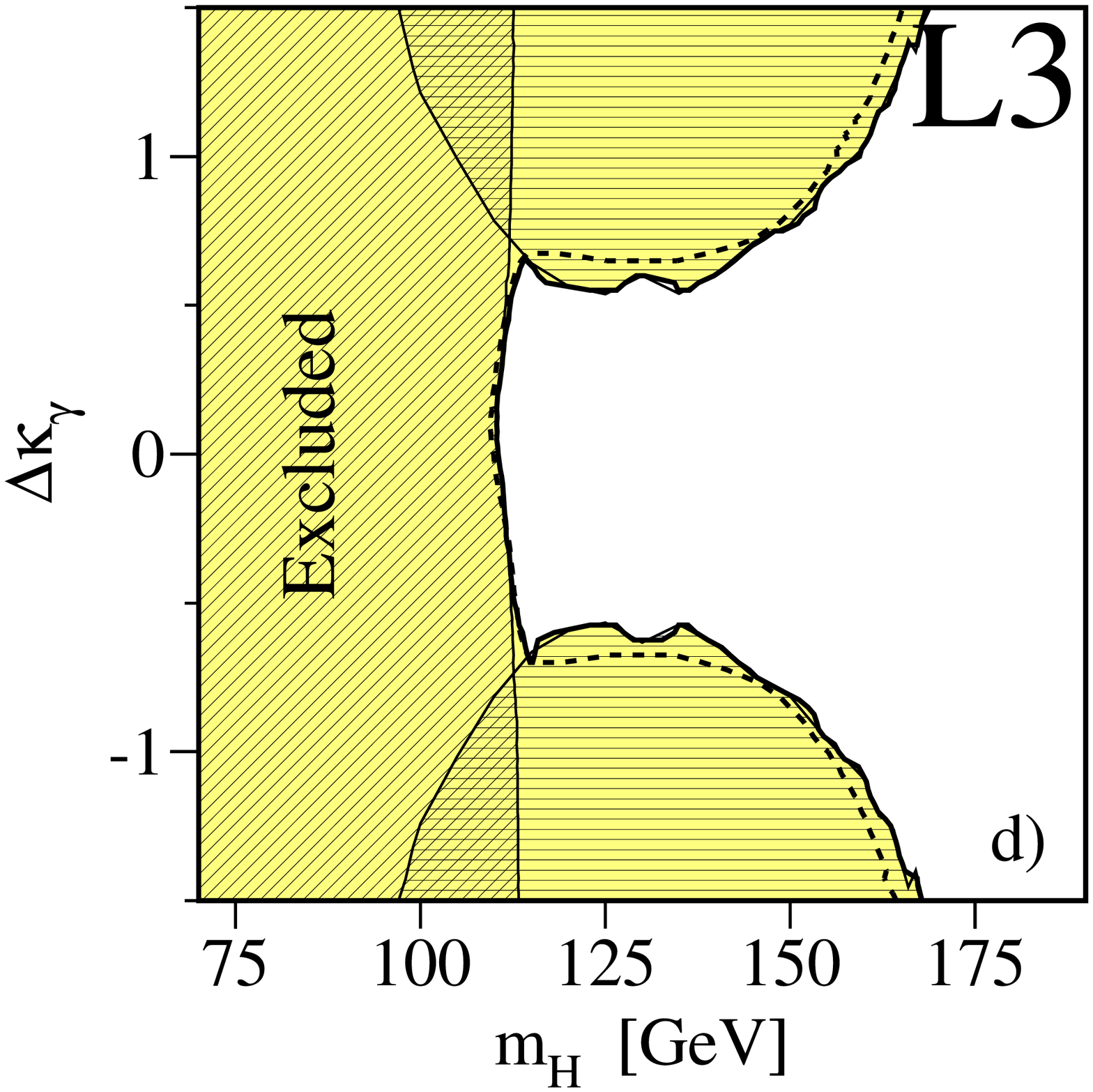} \\
\end{tabular}
\end{center}
\caption{Regions excluded at 95\% CL as a function of the Higgs mass
for the anomalous couplings: a) $\d$, b) $\db$, c) $\dg1z$ and
d) $\dkg$.  The limits on each coupling are obtained under the assumption
that the other three couplings are equal to zero.  
The dashed line indicates the expected limit in the absence of a
signal. The different
hatched regions show the limits obtained by the most sensitive
analyses: $\eeggg$, $\eeeegg$, $\eezgg$, $\eehz$ and $\eewwg$.
}
\label{fig:ac_limits}
\end{figure}

Figure \ref{fig:ac_limits} presents the limits on $\d$, $\db$, $\dg1z$
and $\dkg$ as a function of $\MH$. A coupling at the time is
considered, fixing the others to zero. Limits from the most sensitive
channels are shown in addition to the combined results.

The region $\MH\lesssim\rts-\MZ$ is excluded by the $\eehz$ search
for any value of the four couplings.  The fermiophobic search
$\eehz$, with $\htogg$, is sensitive to large values of $\d$ and
$\db$, for which there is an enhancement of the $\htogg$ branching
fraction. The standard search $\eehz$, with $\htobb$ or $\tautau$, covers
the region $\d\approx\db\approx 0$. 
A region for $\MH \sim 97 \GeV$ in the $\d$ \textit{vs.} $\MH$ plane of Figure \ref{fig:ac_limits}a
is not excluded due to an excess of events observed in the
$\eehz$ search~\cite{l3sm_202}.

The $\eeggg$ and $\eeeegg$ channels have a large sensitivity if
the $\Ho\gamma\gamma$ coupling is large, {\it i.e.} when
$\d\sintw+\db\costw$ has a sizable value (Figures
\ref{fig:ac_limits}a and \ref{fig:ac_limits}b). On the other hand,
the $\eezgg$ process has a dominant role when the channel $\htogg$
is suppressed, which occurs for the couplings $\dg1z$ and $\dkg$
in the mass region $\MZ < \MH < 2\MW$ (Figures
\ref{fig:ac_limits}c and \ref{fig:ac_limits}d).

The contribution from the $\eewwg$ process to the limits presented
in Figures \ref{fig:ac_limits}a and \ref{fig:ac_limits}c is small
and restricted to $\MH \sim 160 \GeV$. This happens since a large
decay width for $\htoww$ corresponds to large values of $\d$ or
$\dg1z$ which also imply large widths for the competing modes
$\htogg$ and $\htozg$.

The sensitivity of the analysis degrades rapidly when $\MH$ approaches
the $2\MW$ threshold, where the $\htogg$ and $\htozg$ are no longer
dominant, even in the presence of relatively large anomalous
couplings.

Several sources of systematic uncertainties are investigated and their
impact on the signal efficiency and 
background level is evaluated. The limited Monte Carlo statistics
affects the signal by less than 2\% and the background by 8\% for the
photonic channels and less than 4\% for the hadronic channels.
The accuracy of the cross section calculation for background processes adds
less than 0.4\% to the uncertainty in the background normalisation.
The systematic uncertainty due to the selection procedure was estimated
by varying the most important selection criteria and was found to be less than 1\%.
In particular, the effect of the limited knowledge of the energy scale
of the electromagnetic calorimeter has a small impact
in the limits.

The combined effect of the systematic uncertainties is included in the limits shown
in Figure~\ref{fig:ac_limits}. It degrades the limits by at most 4\%, slightly depending on the
coupling and the Higgs mass hypothesis.

We verified that possible effects of angular dependence of the
efficiency on the value of the anomalous couplings is negligible for
the  $\eehz$ process. No such effects are expected for the $\eehee$
and $\eehg$ processes.

\subsection{Two-dimensional limits}

\begin{figure}[htbp]
\begin{center}
    \includegraphics[width=10cm]{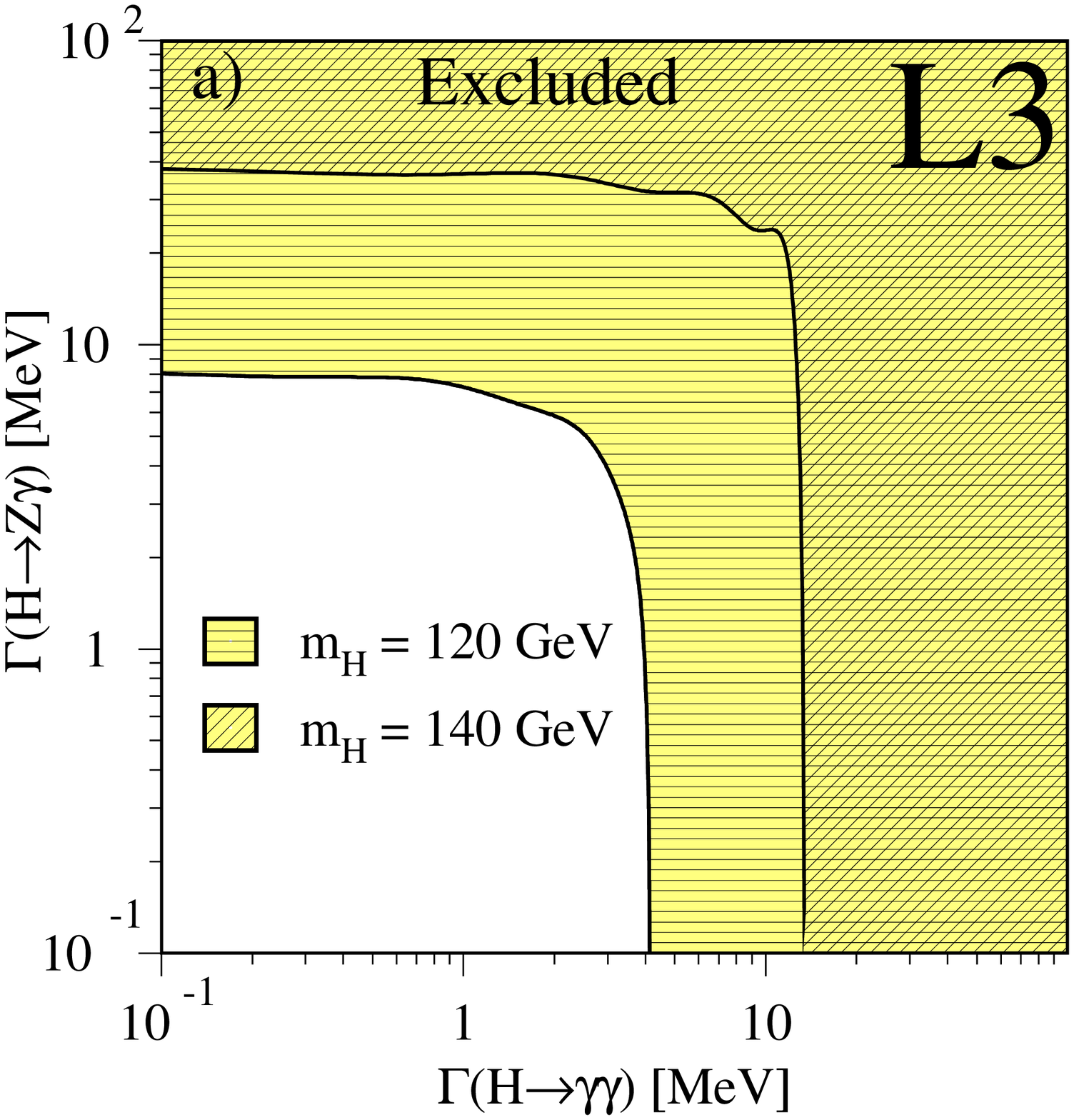}\\
    \includegraphics[width=10cm]{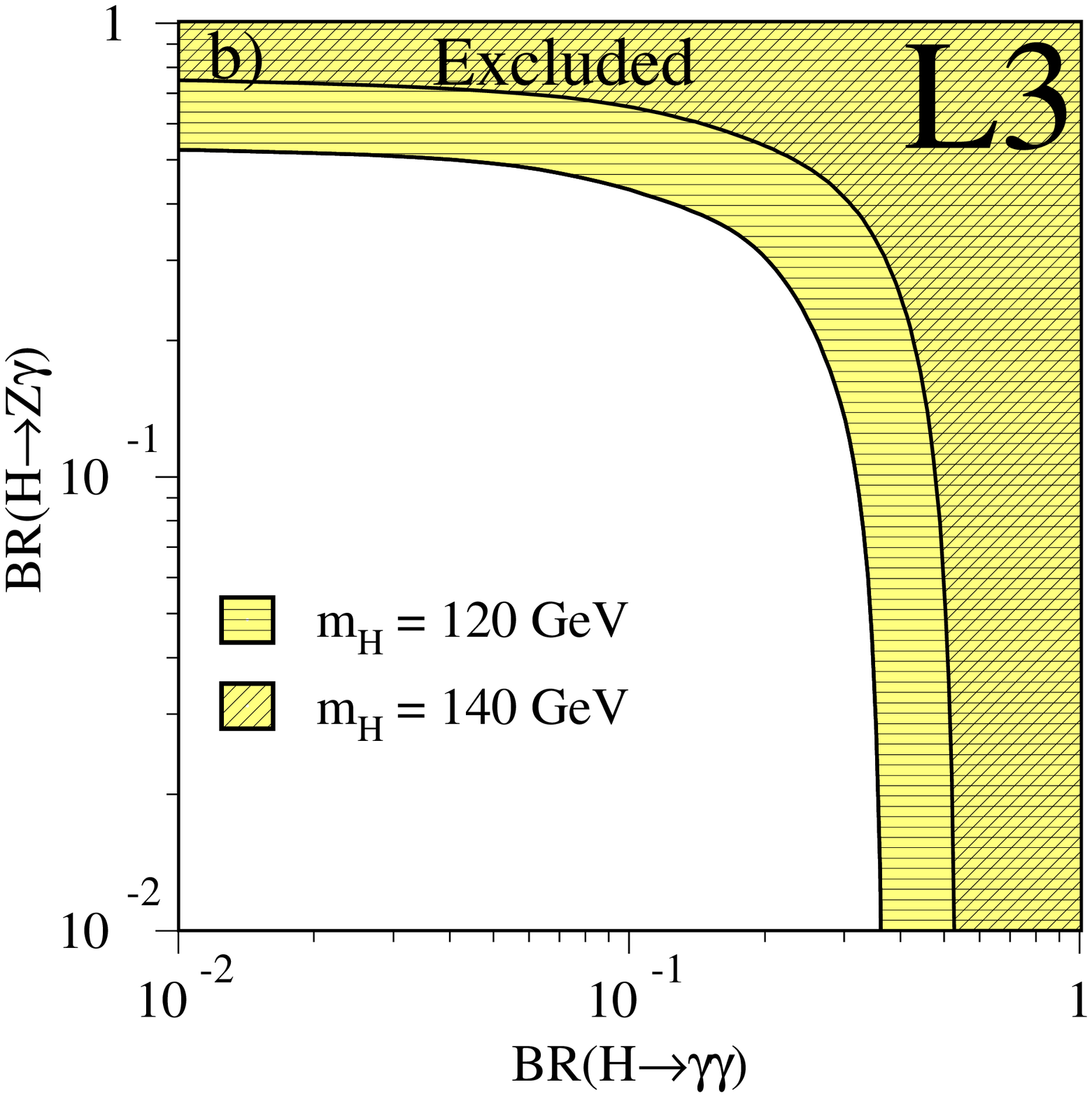}
\end{center}
\caption{Regions excluded at 95\% CL for: a) the partial widths
$\Gamma (\Ho\ra\Zo\gamma)$ {\it vs.}  $\Gamma (\Ho\ra\gamma\gamma)$
and b) the branching fractions $\Brhzg$ {\it vs.} $\Brhgg$ in presence
of the $\d$ and $\db$ anomalous couplings. Two values of the
Higgs boson mass are considered. The results are consistent with the
tree level Standard Model expectations $\Gamma
(\Ho\ra\Zo\gamma)\approx\Gamma (\Ho\ra\gamma\gamma)\approx 0$.}
\label{fig:gg_zg}
\end{figure}

Assuming the absence of large anomalous $\WWZ$ and $\WWG$ couplings, \textit{i.e.}
$\dg1z=\dkg=0$~\cite{l3_tgc}, the $\Ho\gamma\gamma$ and
$\Ho\Zo\gamma$ couplings are parametrized via the following subset of
effective operators:
\begin{eqnarray}
{\cal L}_{\rm eff} = g_{\HGG}~\Ho \A_{\mu\nu} \A^{\mu\nu}
       ~+~g^{(2)}_{\HZG}~\Ho \A_{\mu\nu} \Zo^{\mu\nu}
       ~+~ h.c.
       \label{eq:lagrangian2}
\end{eqnarray}

\noindent where the dependence of $g_{\HGG}$ and $g^{(2)}_{\HZG}$
 on the $\d$ and $\db$ couplings is given by Equations~\ref{eq:gg}
and~\ref{eq:zg}. This Lagrangian is used to compute the maximal
partial widths and branching fractions of the decays
$\Ho\ra\Zo\gamma$ and $\Ho\ra \gamma\gamma$, allowed by the limits
on $\d$ and $\db$. The results are presented in Figure
\ref{fig:gg_zg} for two different Higgs masses, in the region of
interest for Higgs searches at future colliders. The results are consistent with the
tree level Standard Model expectations $\Gamma
(\Ho\ra\Zo\gamma)\approx\Gamma (\Ho\ra\gamma\gamma)\approx 0$.

\bibliographystyle{l3style}

\newpage

\typeout{   }     
\typeout{Using author list for paper 281 -  }
\typeout{$Modified: Jul 15 2001 by smele $}
\typeout{!!!!  This should only be used with document option a4p!!!!}
\typeout{   }
%
%
%
%
%
%

\newcount\tutecount  \tutecount=0
\def\tutenum#1{\global\advance\tutecount by 1 \xdef#1{\the\tutecount}}
\def\tute#1{$^{#1}$}
\tutenum\aachen            
\tutenum\nikhef            
\tutenum\mich              
\tutenum\lapp              
\tutenum\basel             
\tutenum\lsu               
\tutenum\beijing           
\tutenum\bologna           
\tutenum\tata              
\tutenum\ne                
\tutenum\bucharest         
\tutenum\budapest          
\tutenum\mit               
\tutenum\panjab            
\tutenum\debrecen          
\tutenum\dublin            
\tutenum\florence          
\tutenum\cern              
\tutenum\wl                
\tutenum\geneva            
\tutenum\hamburg           
\tutenum\hefei             
\tutenum\lausanne          
\tutenum\lyon              
\tutenum\madrid            
\tutenum\florida           
\tutenum\milan             
\tutenum\moscow            
\tutenum\naples            
\tutenum\cyprus            
\tutenum\nymegen           
\tutenum\caltech           
\tutenum\perugia           
\tutenum\peters            
\tutenum\cmu               
\tutenum\potenza           
\tutenum\prince            
\tutenum\riverside         
\tutenum\rome              
\tutenum\salerno           
\tutenum\ucsd              
\tutenum\sofia             
\tutenum\korea             
\tutenum\taiwan            
\tutenum\tsinghua          
\tutenum\purdue            
\tutenum\psinst            
\tutenum\zeuthen           
\tutenum\eth               

{
\parskip=0pt
\noindent
{\bf The L3 Collaboration:}
\ifx\selectfont\undefined
 \baselineskip=10.8pt
 \baselineskip\baselinestretch\baselineskip
 \normalbaselineskip\baselineskip
 \ixpt
\else
 \fontsize{9}{10.8pt}\selectfont
\fi
\medskip
\tolerance=10000
\hbadness=5000
\raggedright
\hsize=162truemm\hoffset=0mm
\def\r{\rlap,}
\noindent

P.Achard\r\tute\geneva\ 
O.Adriani\r\tute{\florence}\ 
M.Aguilar-Benitez\r\tute\madrid\ 
J.Alcaraz\r\tute{\madrid}\ 
G.Alemanni\r\tute\lausanne\
J.Allaby\r\tute\cern\
A.Aloisio\r\tute\naples\ 
M.G.Alviggi\r\tute\naples\
H.Anderhub\r\tute\eth\ 
V.P.Andreev\r\tute{\lsu,\peters}\
% ...
\newpage
\begin{list}{A}{\itemsep=0pt plus 0pt minus 0pt\parsep=0pt plus 0pt minus 0pt
                \topsep=0pt plus 0pt minus 0pt}
\item[\aachen]
 III. Physikalisches Institut, RWTH, D-52056 Aachen, Germany$^{\S}$
\item[\nikhef] National Institute for High Energy Physics, NIKHEF, 
     and University of Amsterdam, NL-1009 DB Amsterdam, The Netherlands
\item[\mich] University of Michigan, Ann Arbor, MI 48109, USA
\item[\lapp] Laboratoire d'Annecy-le-Vieux de Physique des Particules, 
     LAPP,IN2P3-CNRS, BP 110, F-74941 Annecy-le-Vieux CEDEX, France
\item[\basel] Institute of Physics, University of Basel, CH-4056 Basel,
     Switzerland
\item[\lsu] Louisiana State University, Baton Rouge, LA 70803, USA
\item[\beijing] Institute of High Energy Physics, IHEP, 
  100039 Beijing, China$^{\triangle}$ 
\item[\bologna] University of Bologna and INFN-Sezione di Bologna, 
     I-40126 Bologna, Italy
\item[\tata] Tata Institute of Fundamental Research, Mumbai (Bombay) 400 005, India
\item[\ne] Northeastern University, Boston, MA 02115, USA
\item[\bucharest] Institute of Atomic Physics and University of Bucharest,
     R-76900 Bucharest, Romania
\item[\budapest] Central Research Institute for Physics of the 
     Hungarian Academy of Sciences, H-1525 Budapest 114, Hungary$^{\ddag}$
\item[\mit] Massachusetts Institute of Technology, Cambridge, MA 02139, USA
\item[\panjab] Panjab University, Chandigarh 160 014, India
\item[\debrecen] KLTE-ATOMKI, H-4010 Debrecen, Hungary$^\P$
\item[\dublin] Department of Experimental Physics,
  University College Dublin, Belfield, Dublin 4, Ireland
\item[\florence] INFN Sezione di Firenze and University of Florence, 
     I-50125 Florence, Italy
\item[\cern] European Laboratory for Particle Physics, CERN, 
     CH-1211 Geneva 23, Switzerland
\item[\wl] World Laboratory, FBLJA  Project, CH-1211 Geneva 23, Switzerland
\item[\geneva] University of Geneva, CH-1211 Geneva 4, Switzerland
\item[\hamburg] University of Hamburg, D-22761 Hamburg, Germany
\item[\hefei] Chinese University of Science and Technology, USTC,
      Hefei, Anhui 230 029, China$^{\triangle}$
\item[\lausanne] University of Lausanne, CH-1015 Lausanne, Switzerland
\item[\lyon] Institut de Physique Nucl\'eaire de Lyon, 
     IN2P3-CNRS,Universit\'e Claude Bernard, 
     F-69622 Villeurbanne, France
\item[\madrid] Centro de Investigaciones Energ{\'e}ticas, 
     Medioambientales y Tecnol\'ogicas, CIEMAT, E-28040 Madrid,
     Spain${\flat}$ 
\item[\florida] Florida Institute of Technology, Melbourne, FL 32901, USA
\item[\milan] INFN-Sezione di Milano, I-20133 Milan, Italy
\item[\moscow] Institute of Theoretical and Experimental Physics, ITEP, 
     Moscow, Russia
\item[\naples] INFN-Sezione di Napoli and University of Naples, 
     I-80125 Naples, Italy
\item[\cyprus] Department of Physics, University of Cyprus,
     Nicosia, Cyprus
\item[\nymegen] University of Nijmegen and NIKHEF, 
     NL-6525 ED Nijmegen, The Netherlands
\item[\caltech] California Institute of Technology, Pasadena, CA 91125, USA
\item[\perugia] INFN-Sezione di Perugia and Universit\`a Degli 
     Studi di Perugia, I-06100 Perugia, Italy   
\item[\peters] Nuclear Physics Institute, St. Petersburg, Russia
\item[\cmu] Carnegie Mellon University, Pittsburgh, PA 15213, USA
\item[\potenza] INFN-Sezione di Napoli and University of Potenza, 
     I-85100 Potenza, Italy
\item[\prince] Princeton University, Princeton, NJ 08544, USA
\item[\riverside] University of Californa, Riverside, CA 92521, USA
\item[\rome] INFN-Sezione di Roma and University of Rome, ``La Sapienza",
     I-00185 Rome, Italy
\item[\salerno] University and INFN, Salerno, I-84100 Salerno, Italy
\item[\ucsd] University of California, San Diego, CA 92093, USA
\item[\sofia] Bulgarian Academy of Sciences, Central Lab.~of 
     Mechatronics and Instrumentation, BU-1113 Sofia, Bulgaria
\item[\korea]  The Center for High Energy Physics, 
     Kyungpook National University, 702-701 Taegu, Republic of Korea
\item[\taiwan] National Central University, Chung-Li, Taiwan, China
\item[\tsinghua] Department of Physics, National Tsing Hua University,
      Taiwan, China
\item[\purdue] Purdue University, West Lafayette, IN 47907, USA
\item[\psinst] Paul Scherrer Institut, PSI, CH-5232 Villigen, Switzerland
\item[\zeuthen] DESY, D-15738 Zeuthen, Germany
\item[\eth] Eidgen\"ossische Technische Hochschule, ETH Z\"urich,
     CH-8093 Z\"urich, Switzerland
\item[\S]  Supported by the German Bundesministerium 
        f\"ur Bildung, Wissenschaft, Forschung und Technologie.
\item[\ddag] Supported by the Hungarian OTKA fund under contract
numbers T019181, F023259 and T037350.
\item[\P] Also supported by the Hungarian OTKA fund under contract
  number T026178.
\item[$\flat$] Supported also by the Comisi\'on Interministerial de Ciencia y 
        Tecnolog{\'\i}a.
\item[$\sharp$] Also supported by CONICET and Universidad Nacional de La Plata,
        CC 67, 1900 La Plata, Argentina.
\item[$\triangle$] Supported by the National Natural Science
  Foundation of China.
\end{list}
}
\vfill


\newpage

\renewcommand\arraystretch{1.3}

\begin{table}[htbp]
  \begin{center}
    \begin{tabular}{|rcl|c|c|c|c|}
      \hline
      \multicolumn{3}{|c|}{Production}  & \multicolumn{4}{|c|}{Decay mode} \\ \cline{4-7}
      \multicolumn{3}{|c|}{mechanism}   & $\htogg$ & $\htozg$ & $\htoww$ & $\htoff$ \\
      \hline
$\rm e^+e^-$ & $\rightarrow$ & $\rm H\gamma$
 & 3 $\gamma$ & 2 $\gamma$ + 2 jets &  1 $\gamma$ + 4 jets & 1 $\gamma + \bbbar$ \cite{L3_anom}\\
$\rm e^+e^-$ & $\rightarrow$ & $\rm He^+e^-$
&  2 $\gamma$ + $p$\hspace{-.18cm}/\hspace{+.01cm} &   --     &  --   & $\bbbar$ + $p$\hspace{-.18cm}/\hspace{+.01cm} \cite{L3_anom} \\
$\rm e^+e^-$ & $\rightarrow$ & $\rm HZ$
& 2 $\gamma + \ffbar$ \cite{l3aura_paper}&   --     &  --   & $\ffbar\fpfpbar$ \cite{l3sm_paper} \\ \hline
    \end{tabular}
  \end{center}
  \caption{Experimental signatures for the search for anomalous
    couplings in the Higgs sector. The symbol
    $p$\hspace{-.18cm}/\hspace{+.01cm} denotes missing energy and
    momentum. Searches in the $\eebbg$ and
    $\eeeebb$ channels are only performed at $\rts=189\GeV$~\protect\cite{L3_anom}.}
  \label{tab:signatures}
\end{table}

\renewcommand\arraystretch{1.0}

\begin{table}[htbp]
\begin{center}
 \begin{tabular}{|l|rrrrrrr|} \hline
 $\rts$ (\GeV)              &  188.6 & 191.6 & 195.5 &199.5 &201.7 &204.8 & 206.6 \\
 $\cal{L}$ ($\rm{pb^{-1}}$) &  176.8 &  28.8 &  82.4 & 67.6 & 36.1 & 74.7 & 135.6 \\ \hline
\end{tabular}
\end{center}
\caption{Average centre-of-mass energy and integrated luminosity of the
 data samples used for the search for anomalous
    couplings in the Higgs sector.}
\label{tab:energies}
\end{table}

\begin{table}[htbp]
\begin{center}
 \begin{tabular}{|c|ccc|ccc|ccc|ccc|} \cline{2-13}
             \multicolumn{1}{c|}{}
           & \multicolumn{12}{|c|}{$\ee \ra$} \\ \hline
           & \multicolumn{3}{|c|}{$\Ho\gamma \ra \gamma\gamma\gamma$}
           & \multicolumn{3}{|c|}{$\ee\Ho    \ra \ee\gamma\gamma   $}
           & \multicolumn{3}{|c|}{$\Ho\gamma \ra \Zo\gamma\gamma   $}
           & \multicolumn{3}{|c|}{$\Ho\gamma \ra \WWstar\gamma     $} \\  \cline{2-13}
       $\MH$  & $N_D$ & $N_B$ & $\epsilon (\%)$ &
                $N_D$ & $N_B$ & $\epsilon (\%)$ &
                $N_D$ & $N_B$ & $\epsilon (\%)$ &
                $N_D$ & $N_B$ & $\epsilon (\%)$ \\ \hline
  70   & 1 & 3.5 & 23.4 & \phantom{0}0 & \phantom{0}0.0  & 19.5 &  --  &  --  & -- & -- &  --  & -- \\
  90   & 2 & 2.7 & 25.8 & \phantom{0}6 & \phantom{0}1.7  & 24.2 &  --  &  --  & -- & -- &  --  & -- \\
  110  & 3 & 3.1 & 26.9 & \phantom{0}9 & \phantom{0}4.9  & 28.5 &  68  & \phantom{0}72.8 & 22.7 & -- &  --  & -- \\
  130  & 2 & 2.4 & 28.7 & 11 & 10.9 & 30.4 &  15  & \phantom{0}18.2 & 22.4 & \phantom{0}10 & \phantom{0}11.5 & 18.0 \\
  150  & 4 & 4.0 & 28.8 & 19 & 19.9 & 31.9 &  \phantom{0}9  & \phantom{0}14.4 & 24.1 & \phantom{0}22 & \phantom{0}22.8 & 25.5   \\
  170  & 9 & 9.3 & 28.2 & 38 & 49.7 & 32.4 &  31  & \phantom{0}41.0 & 25.6 & \phantom{0}72 & \phantom{0}74.7 & 26.8   \\
  190  & 3 & 8.9 & 22.9 & 24 & 29.5 & 30.1 &  96  &101.0 & 22.5 &113 & 107.3 & 19.5  \\ \hline
\end{tabular}
\end{center}
\caption{Numbers of observed, $N_D$, and expected, $N_B$, events and
signal selection efficiencies, $\epsilon$, for different analysis
channels and values of the Higgs mass.  Centre-of-mass energies in the
range $189 \GeV < \rts < 209 \GeV$ are considered for the $\eeggg$
and $\eeeegg$ channels, while the $\eezgg$ and $\eewwg$ channels
are analysed in the $192 \GeV < \rts < 209 \GeV$ range.}
\label{tab:events}
\end{table}

\newpage

\end{document}